%% file: CFT-09-012_temp.tex
\pdfoutput=1
\documentclass[11pt,twoside,a4paper,pdftex,cmspaper,final,collab]{cms-tdr}

\begin{document}\cmsNoteHeader{CFT-09-012}
%
%
%

%
%
\hyphenation{env-iron-men-tal}
\hyphenation{had-ron-i-za-tion}
\hyphenation{cal-or-i-me-ter}
\hyphenation{de-vices}
%
%
\RCS$Revision: 1.13 $
\RCS$Date: 2010/01/15 08:46:04 $
\RCS$Name:  $
\input{ptdr-definitions}
\cmsNoteHeader{09-012}
\title{Performance of the CMS Drift Tube Chambers \\
  with Cosmic Rays }


\date{\today}

\abstract{
Studies of the performance of the CMS drift tube
barrel muon system are described, with results based
on data collected during the CMS Cosmic Run at Four
Tesla.  For most of these data, the solenoidal
magnet was operated with a central field of 3.8 T.
The analysis of data from 246 out of a total of
250 chambers indicates a very good muon reconstruction
capability, with a coordinate resolution for a single
hit of about 260~$\mu$m, and a nearly $100\%$ efficiency for the drift tube
cells. The resolution of the track
direction measured in the bending plane is about $1.8$~mrad,
and the efficiency to reconstruct a segment in a single
chamber is higher than $99\%$. The CMS simulation of cosmic rays reproduces well
the performance of the barrel muon detector.
}

\hypersetup{%
pdfauthor={Ugo Gasparini},%
pdftitle={Performance of the CMS Drift Tube Chambers with Cosmic Rays},%
pdfsubject={CMS},%
pdfkeywords={CMS, physics, software, muons}}

\maketitle 

\section{Introduction}

The primary goal of the Compact Muon Solenoid (CMS) experiment~\cite{CMSexp}
is to explore particle physics at the TeV energy scale, exploiting the
proton-proton collisions delivered by the Large Hadron Collider (LHC) at CERN.
The central feature of the Compact Muon Solenoid apparatus is a
superconducting solenoid, of $6$~m internal diameter, providing a field of
3.8~T. Within the field volume are the silicon pixel and strip tracking detectors, the
crystal electromagnetic calorimeter and the brass/scintillator hadron
calorimeter.  Muons are measured in gas-ionization detectors embedded
in the steel return yoke.  In addition to the barrel and endcap detectors,
CMS has extensive forward calorimetry.

In autumn of 2008, after closing the CMS detector in preparation for the LHC
start-up and the first underground test of the magnet, CMS undertook a long period (about 1 month) of data taking,
collecting about 270 million cosmic ray events with varying detector and trigger conditions.
 Data were collected both without and with magnetic field (at various values of the current in the coil of the
 solenoid). In this ``Cosmic Run At Four Tesla'' (CRAFT), the large majority of
 the data were collected with a magnetic field of $B$ = 3.8~T in the volume of the solenoid.
 Almost all CMS sub-detectors were active and included in the data acquisition~\cite{CRAFTgen}.

 In summer 2006, cosmic ray data were taken on the surface with the detector closed,
 the ``Magnet Test and Cosmic Challenge'' (MTCC)~\cite{MTCC}.
 In that period only a small part (about $5\%$) of the muon detector was equipped for readout, and
 the tracking detectors were not installed inside the coil.
 Many results on the muon detector performance~\cite{MTCCvdrift} and measurements of physical quantities related to the
 cosmic ray properties~\cite{muratio} were obtained. The CRAFT exercise allowed the extension
 of those studies of muon reconstruction and identification to the
 entire system, and in much greater detail.

 This paper addresses muon reconstruction in the drift tube chambers of the barrel muon system,
 hereafter referred to as ``DT chambers'', focusing on the reconstruction of local hits and track segments
 in the chambers. Information from this reconstruction, together with the output of the local
 reconstruction of other CMS subsystems, is used as input to the following stage of the global muon
 reconstruction~\cite{CFTmureco}.
Detailed comparisons of different track segments belonging to the same track, but measured
in different stations, were performed, using in addition information from the internal tracking devices.
The non-bunched structure of the cosmic rays affects the time measurements in the DT cells
and hence the position resolution obtained in the initial stage of the reconstruction process. Despite this, and
  the fact that cosmic rays illuminate a large part of the detector  quite differently
  from the muons produced in proton-proton collisions, it is shown that the final reconstruction
  performance is very good, not far from the performance expected from test beam studies and
  required for operation at the LHC.

 The muon barrel system and its operating trigger conditions are described in Section~2.
 After a brief discussion of the Monte Carlo simulation of cosmic ray data in Section~3,
 the main features of the local muon reconstruction in the DT chambers
 are summarized in Section~4. The results on hit reconstruction and local track segments are given
 in Sections~5 and 6, respectively.

\section{DT Chamber Setup and Trigger Conditions}

\begin{figure}[htbp]
 \begin{center}
  \resizebox{14cm}{!}{\includegraphics{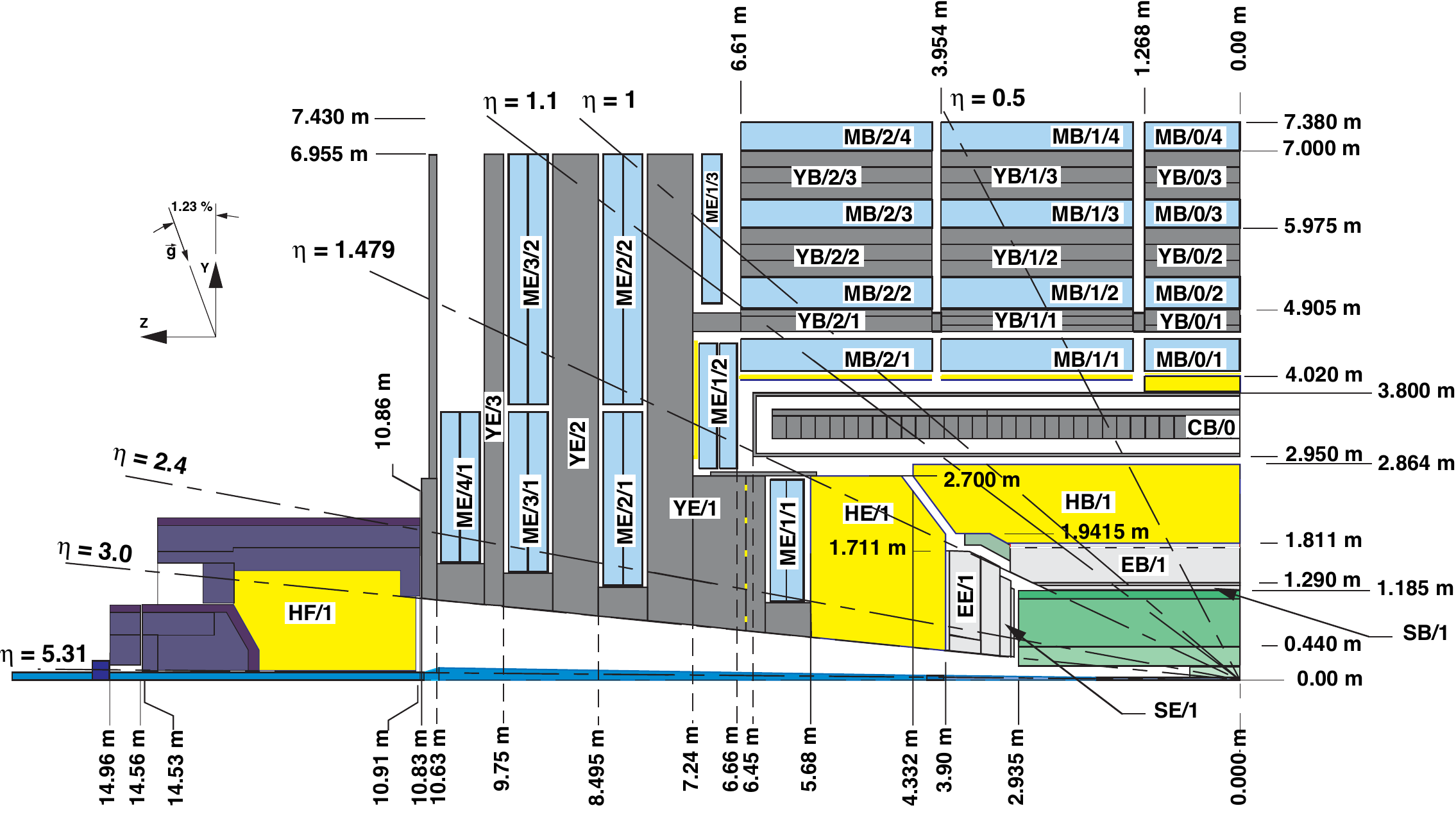}} \\
  \resizebox{14cm}{!}{\includegraphics{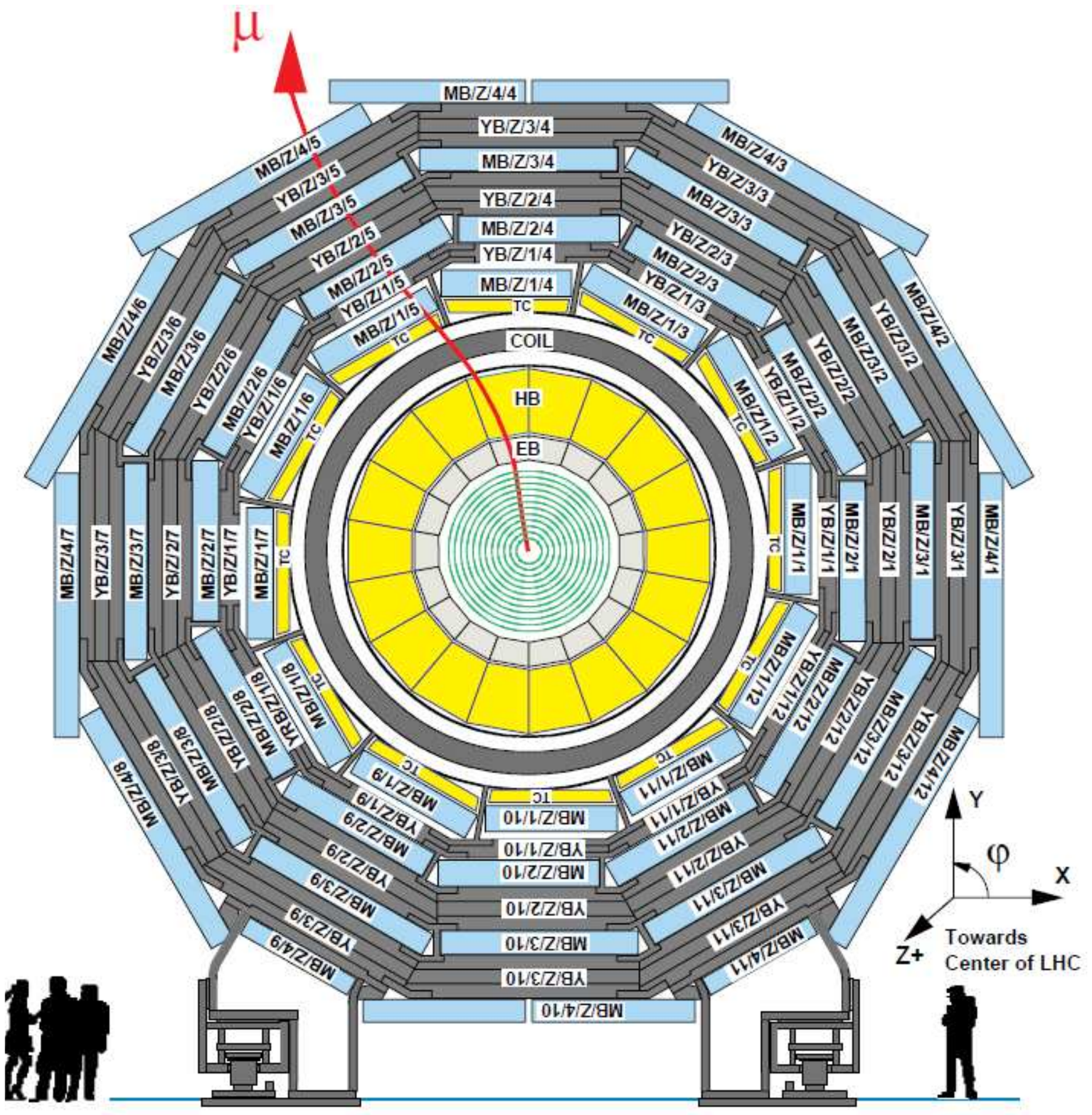}}
  \caption{ Schematic view of the CMS detector. Top: longitudinal view of one quarter of the detector.
  Bottom: transverse view at $z=0$. The barrel muon detector elements are denoted as MBZ/N/S, where Z=$-2$,...$+2$ is
  the barrel wheel number, N=1...4 the station number and S=1...12 the sector number. Similarly, the steel return yokes
  are denoted YBZ/N/S.}
  \label{CMSview}
 \end{center}
\end{figure}

A schematic view of CMS is shown in Fig.~\ref{CMSview}. As seen in the longitudinal view, the barrel part of the detector
is divided in 5 wheels, named YB$0$, YB$\pm1$, YB$\pm2$ throughout this paper. All 250 DT chambers of the barrel muon
system~\cite{MUtdr} were installed in the wheels and equipped for data taking at beginning of
CRAFT. Two chambers were subsequently switched off for most of the data acquisition period due to hardware problems,
which were solved by interventions carried out in the winter 2009 shutdown.
Each wheel is divided into 12 sectors, each covering an azimuthal region of 30 degrees.
Sectors are numbered anticlockwise, starting from the right-most vertical sector shown in
Fig.~\ref{CMSview} (bottom) in the direction of increasing azimuthal angle,~$\phi$.
There are four layers of chambers (stations), named MB1-MB4 starting from the innermost one.
In each station there is one DT chamber per sector, except
in the uppermost (lowermost) sector, named sector 4 (sector 10), where the station MB4
is physically made of two DT chambers.

There is a vertical shaft leading from the cavern to the surface originally used for lowering parts
of the CMS detector into the cavern. This shaft is located on the negative $z$ side of the detector, and
as a consequence, the cosmic rays flux was not uniform along the $z$ coordinate of CMS, decreasing by  about $20\%$  when passing
from wheel YB$-2$ to YB$2$.

\begin{figure}[htbp]
 \begin{center}
  \resizebox{12cm}{!}{\includegraphics{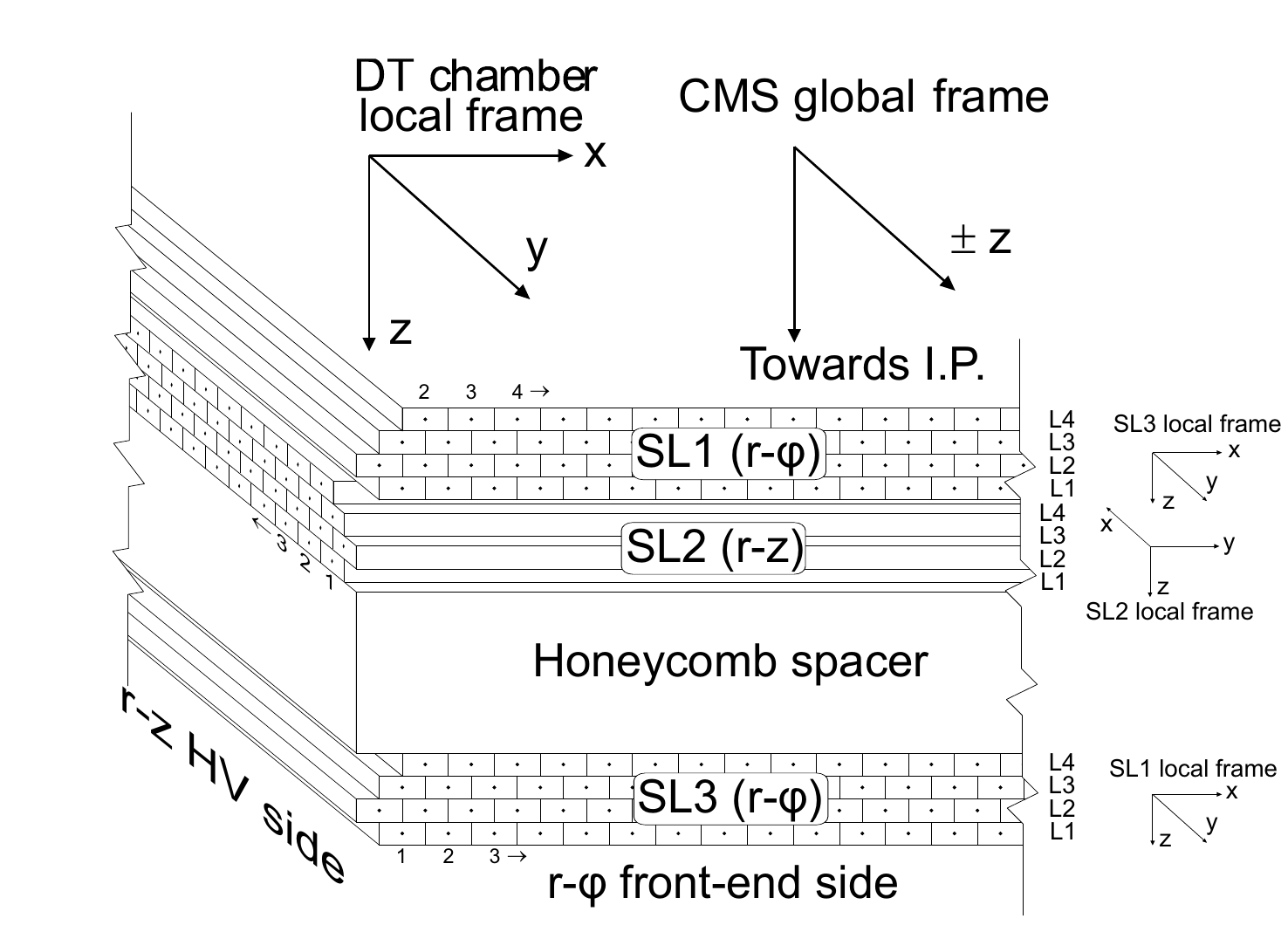}} \\
  \vspace{2cm}
 \resizebox{10cm}{!}{\includegraphics{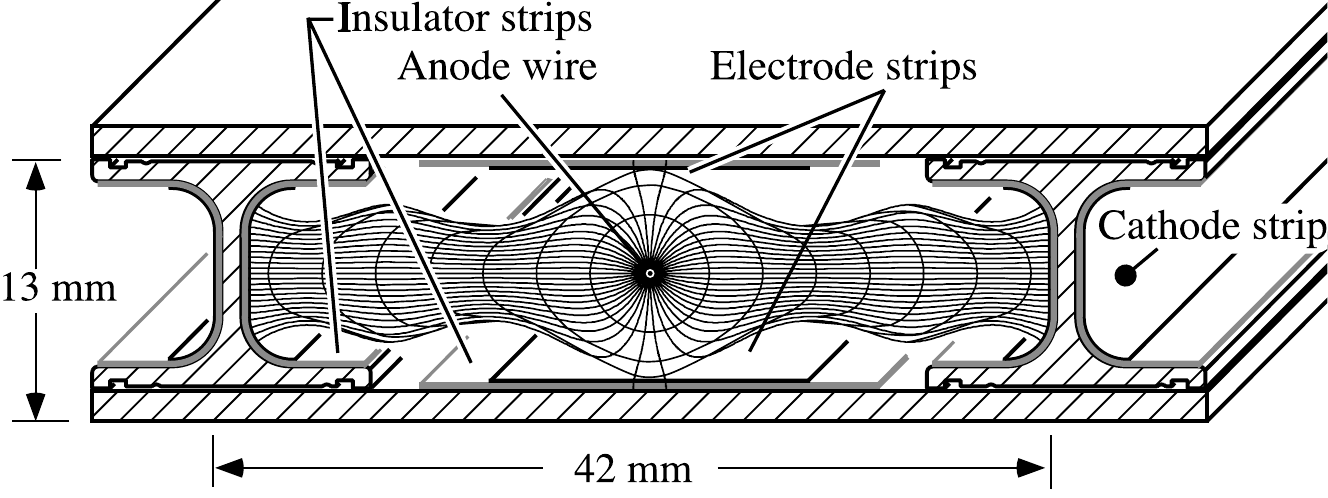}}
 \vspace{1cm}
  \caption{ Top: schematic layout of a DT chamber. The distance between the
  innermost and outermost superlayer (SL) in the chamber is about 25 cm. The SL1 and SL3 superlayers measure
  the  $r$-$\phi$ coordinate in the bending plane of CMS;
  the  SL2 superlayer measures the $z$ coordinate, along the direction parallel to the beam (perpendicular to the plane of the figure).
  Bottom: layout of a DT cell, showing the electric field lines in the gas volume.}
  \label{DTchamber}
 \end{center}
\end{figure}

A schematic layout of a DT chamber and of a DT cell are shown in
Fig.~\ref{DTchamber}. In each chamber there are 12 layers of contiguous drift tube cells grouped in three
``superlayers'' (SL) with 4 staggered layers each;
the innermost and outermost SLs, labeled SL1 and SL3 in the figure, are dedicated to
coordinate measurement in the CMS bending plane
($r$-$\phi$ plane), while in the central SL, labeled SL2, the hits are measured
along the beam axis ($r$-$z$ plane). The outermost stations, named MB4, located outside the steel return yokes of the CMS magnet, have only the two SLs measuring the hit
position in the  $r$-$\phi$ plane. The distance between the anode wires of consecutive cells is $4.2$~cm;
the cells are separated by $1$~mm thick
aluminium I-beams glued between two $2.5$~mm thick aluminium plates separating consecutive layers.
Also visible are the aluminium strips, named ``electrodes'' in the figure, below and above the anode wire of
the cell, which are needed to shape the electric field lines. This field shaping guarantees a good linearity of the cell behaviour
over almost the entire drift volume~\cite{DTtest}. The chambers are operated with an Ar/CO$_2(85/15\%)$ gas mixture. The voltages applied to the electrodes are
$+3600$~V for wires, $+1800$~V for strips, and $-1200$~V for cathodes. The electron drift velocity is about $54$ $\mu$m/ns.
The DT readout electronics is capable of recording multiple hits in the same cell, with a dead time of
$150$~ns between consecutive signals.

 At the operating value of $B=3.8$~T for the magnetic field inside the solenoid,
 typical values of the magnetic field inside the steel return yokes of the magnet structure, where the muon chambers are located,
 range between $1.2$ and $1.8$~T. In the active volume of the DT chambers, the residual magnetic field is generally small (below $0.2$ T),
 except for the innermost chambers in the outermost wheels YB$\pm$2.

The DT chamber Local Trigger~\cite{DTtrigg} performs a rough track reconstruction within each SL and
uniquely assigns the parent bunch crossing number to a track candidate. A Track Correlator processor
associates track segments in the same chamber by combining the information from the SLs of the
$r$-$\phi$ view, enhancing the angular resolution and providing a quality hierarchy of the trigger primitives.
Up to two local trigger primitives are transmitted to the Regional Muon Trigger, which constitutes the
following step of the level-1 muon trigger, running an algorithm called DT TrackFinder. This
algorithm links the track primitives and forms muon candidates, assigning  their angular coordinates
and transverse momentum measurement.
The DT local trigger was operating in all the sectors and wheels of the barrel muon system.
After proper chamber synchronization within the same sector and between neighbouring sectors, the
DT TrackFinder trigger provided a stable cosmic muon rate of about 240 Hz for the entire
one month period of data taking~\cite{DTtriggCRAFT}.
It was operated with an open look-up table configuration
requiring the coincidence of local triggers from at least two chambers in the same sector,
with no requirements on the muon candidate direction and transverse momentum.
The combination of the two chambers used correlated trigger candidates from the trigger processor in
each station, which combines the trigger primitives between the chambers'
SLs in the $r$-$\phi$ bending plane~\cite{DTtriggCRAFT}.

\section{Monte Carlo Simulation of Cosmic Ray Data}

 A simulation of the cosmic muon spectrum~\cite{CosmicsMC}
 has been used to compare the detector performance
 in the simulation to the data. About 20 million events
 with a muon momentum above $4$~ GeV/c, as defined
 on a cylindrical surface of $8$ m radius co-axial with the CMS $z$-axis,
 were generated and processed through the full CMS simulation
 and reconstruction chain. The magnetic field inside the CMS solenoid was set to $B=3.8$~T.
  The muon crossing time at the top of the CMS detector was generated according to a flat distribution within
a $\pm 12.5$~ns time window, to replicate the random arrival
 time  of the muon in a bunch crossing window ($25$~ns) of the trigger. The time
 signals that constitute the Time-to-Digital Converters (TDCs) raw data were generated by the digitization
 algorithm based on the parameterization of the DT cell response described in Ref.~\cite{DTdigi} and tuned on test beam data,
 taking into account the muon time of flight from chamber to chamber.

A realistic representation of misalignments based on the analysis of CRAFT data~\cite{CFTalisoftw}
was implemented in the CMS detector simulation. 
The CMS alignment strategy combines precise survey and photogrammetry information, 
measurements from an optical based muon alignment system~\cite{CFTalihardw}, and the result of the 
alignment procedures based on muon tracks~\cite{CFTalisoftw}. A complete alignment of all muon 
chambers was not available for CRAFT. For the internal geometry of the DT chambers, 
which is relevant for the local reconstruction of the muon tracks,
the spread of the measurements of the layer relative positions measured during chamber 
construction and of the photogrammetry measurements made on reflective targets on the 
exterior of the superlayers were taken into account in the geometrical database of the 
detector. In the simulation, typical RMS deviations from the ideal detector geometry 
are taken to be $100$~$\mu$m, with $30$ - - $40$~$\mu$m systematical uncertainty for the layer 
position, and about $200$~$\mu$m for the superlayer positions inside the chamber.
The positions of the muon chambers in the global CMS reference  system were misaligned 
with a 2 mm Gaussian smearing in x , 4 mm in y and z, reflecting the initial 
uncertainty expected from the available photogrammetry measurements, taken with the 
CMS detector open. The orientations of the chambers in $r-\phi$ and $r-z$ planes were 
smeared by 2~mrad.

\section{Local Reconstruction of Muon Tracks}

In the first stage of the local reconstruction, the hits in each DT cell are reconstructed starting from
the measured  time associated to them, as recorded by the TDCs.
The electron drift time, $t_{\mathrm{drift}}$, is computed from the TDC raw data by performing the following operations:
\begin{itemize}
\item subtraction of the inter-channel synchronization constants, $T_0$s, which correct for different
signal path lengths of readout electronics in the chamber front-end. The $T_0$s are measured using
electronic test pulse signals~\cite{DTcalib}.
\item  subtraction of the ``time-pedestal'', $t_{\mathrm{trig}}$, computed at the superlayer level in each chamber.
The quantity $t_{\mathrm{trig}}$ accounts for the time latency of the Level-1
trigger and the time of flight of the muon to the chamber. It is computed by a calibration procedure that fits
the rising edge of the distribution of the TDC recorded times for
all the cells in the superlayer, as described in detail in Ref.~\cite{DTcalib}.
\end{itemize}

\begin{figure}[htbp]
 \begin{center}
 \resizebox{10cm}{!}{\includegraphics{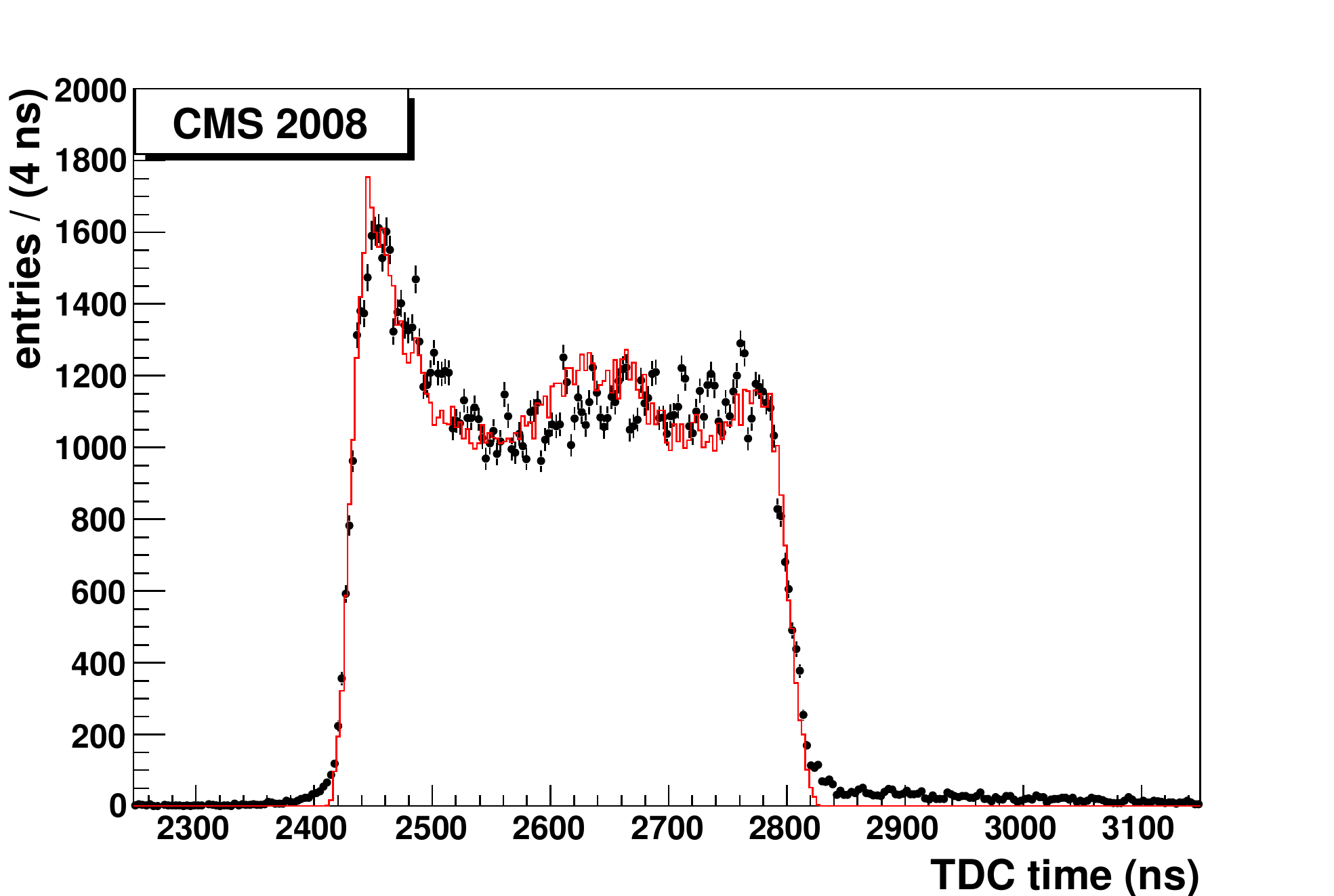}}
  \caption{ Distribution of the signal arrival time in CRAFT (points) and simulated data (full line histogram).
  The arrival time in all the cells from a single superlayer in a chamber are shown,
   after the cell-by-cell equalization based on electronic test-pulse calibration.}
  \label{Tbox}
 \end{center}
\end{figure}

A typical distribution is shown in Fig.~\ref{Tbox} for real and simulated data,
after the measured $T_0$'s have been subtracted cell-by-cell.
The peak at the beginning of the time distribution is due to non-linear effects in the avalanche region
very near (a few wire diameters wide) the anode wire, and to
the occurrence of $\delta$-ray electrons which pass closer the anode
wire than the muon track. The tail in the real data after the ``time-box'' distribution
(i.e. for TDC time greater than $2800$~ns which, for the specific superlayer
shown in the figure, corresponds to the maximum drift length in the cell) is due to ``feed-back'' electrons. These are electrons extracted either from
the cell I-beam or from the aluminium strips (see Fig.~\ref{DTchamber}) by photons produced in the cascade
process initiated by the primary electrons very near the anode wire (these photons are not further considered in the simulation).
The arrival time of the signal associated with these feed-back electrons thus exceeds the maximum drift time in a cell.
The stability of the calibration results and their dependence on trigger conditions and chamber
locations is discussed in Ref.~\cite{DTcalibCRAFT}.

Hits with $t_{\mathrm{drift}}< -3$~ns are discarded, while hits having  $-3<t_{\mathrm{drift}}< 0$~ns are
retained and assigned the position $x=0$ in the
local reference frame of the cell, corresponding to the anode wire position.
The conversion from time measurements to hit positions  in a DT cell~\cite{LocalRecoNote}, leading to
one-dimensional reconstructed hits, or ``rechits'', was performed assuming a constant effective drift velocity in
the whole chamber volume, independent of track position and inclination. This assumption is justified for all chambers
except the innermost stations, MB1$n$ ($n=1...12$), of those mounted on the
YB$2$ and YB$-2$ wheels~\cite{DTcalibCRAFT}. More sophisticated
algorithms~\cite{LocalRecoNote} based on a detailed parametrization of the DT cell behaviour,
developed using simulated data, are currently under study. For the purposes of the present studies, however,
including the MB1 chambers in the outermost wheels, the current algorithm is adequate (once the correct
average value of the drift velocity in these chambers is properly taken into account), as will be shown in Section 5.

For each TDC signal there are two possible rechits due to
the left-right ambiguity on the position with respect to the anode wire inside the cell. This ambiguity is resolved at the track segment building stage~\cite{LocalRecoNote}
by the local pattern recognition algorithm that takes the rechits as input, thanks to the staggered structure
of the cells in the chamber SLs as shown in Fig.~\ref{DTchamber}.
The pattern recognition is initiated by considering all possible pairs of hits (seeds) in different layers,
starting from
the most separated hits in the chamber.
For each seed, additional hits are searched for in all layers and
included in the segment candidate if they are compatible with the extrapolation from the seed within
a loose requirement ($2$~mm). Segment candidates are built by performing a straight-line fit to the
associated hits and sorted on the basis of their total number of hits and $\chi^2$, defined as the sum
of the squares of the hit residuals divided by the hit position error, normalized to the number of degrees of
freedom. The sagitta of the muon track in the (generally small) residual magnetic field in the chamber
volume is negligible.
For each seed, only the segment candidate with the maximum number of hits is considered;
among the candidates with the same number of hits, the one with best $\chi^2$ is selected.
Segments with at least three hits and $\chi^2/NDOF < 20$ are finally retained.

The pattern recognition is performed independently in the $r$-$\phi$ and $r$-$z$ SLs of each chamber
to deliver the so-called 2-dimensional (2D) track segments in both views.
The 2D segments are then paired using all possible combinations to form 4-dimensional
(4D) segments in the chamber, carrying 3-dimensional spatial information and the fitted value of
the arrival time of the muon in the chamber (see next section).
The arrival time of the TDC signal determining the position in a given direction is corrected
for the signal propagation time along the cell wire, using the position information of the associated
hits measured in the orthogonal view of the chamber,
and the rechit position is updated in the 4D segment accordingly.
The 4D segments are used as input to the subsequent stage of the global muon reconstruction that links the
information from different muon stations and from the tracker detector to fit a unique track.
The reconstruction used the standard CMS reconstruction code
that takes into account the alignment corrections obtained from the knowledge of the internal structure
of most chambers, but not yet the complete information of the chambers' position in the CMS structure.

\section{Reconstructed Hits in DT Chambers}

 One-dimensional reconstructed hits in the DT cell are the basic objects
 from which the muon track reconstruction is initiated. This section summarizes the main results concerning the hit resolution
 and reconstruction efficiency.

\subsection{Spatial resolution}

The one-dimensional hits are first determined assuming a fixed arrival time in the chamber of the cosmic muon, $t_0=0$,
inside the $25$ ns wide window associated with the L1 trigger. At this stage the hit resolution is
about $660$ $\mu$m, largely dominated by the uncertainty on $t_0$. Once the local pattern recognition is performed and local segments are built, a re-fit is
performed treating $t_0$ as a free parameter,
recomputing the hit positions and the final segment position and direction. At this final stage of the local
reconstruction, the resolution is about $260$ $\mu$m, in good agreement with the requirements for
collision data~\cite{MUtdr} and the results from test beam measurements~\cite{DTtest}.

\begin{figure}[htbp]
 \begin{center}
  \resizebox{17cm}{!}{\includegraphics{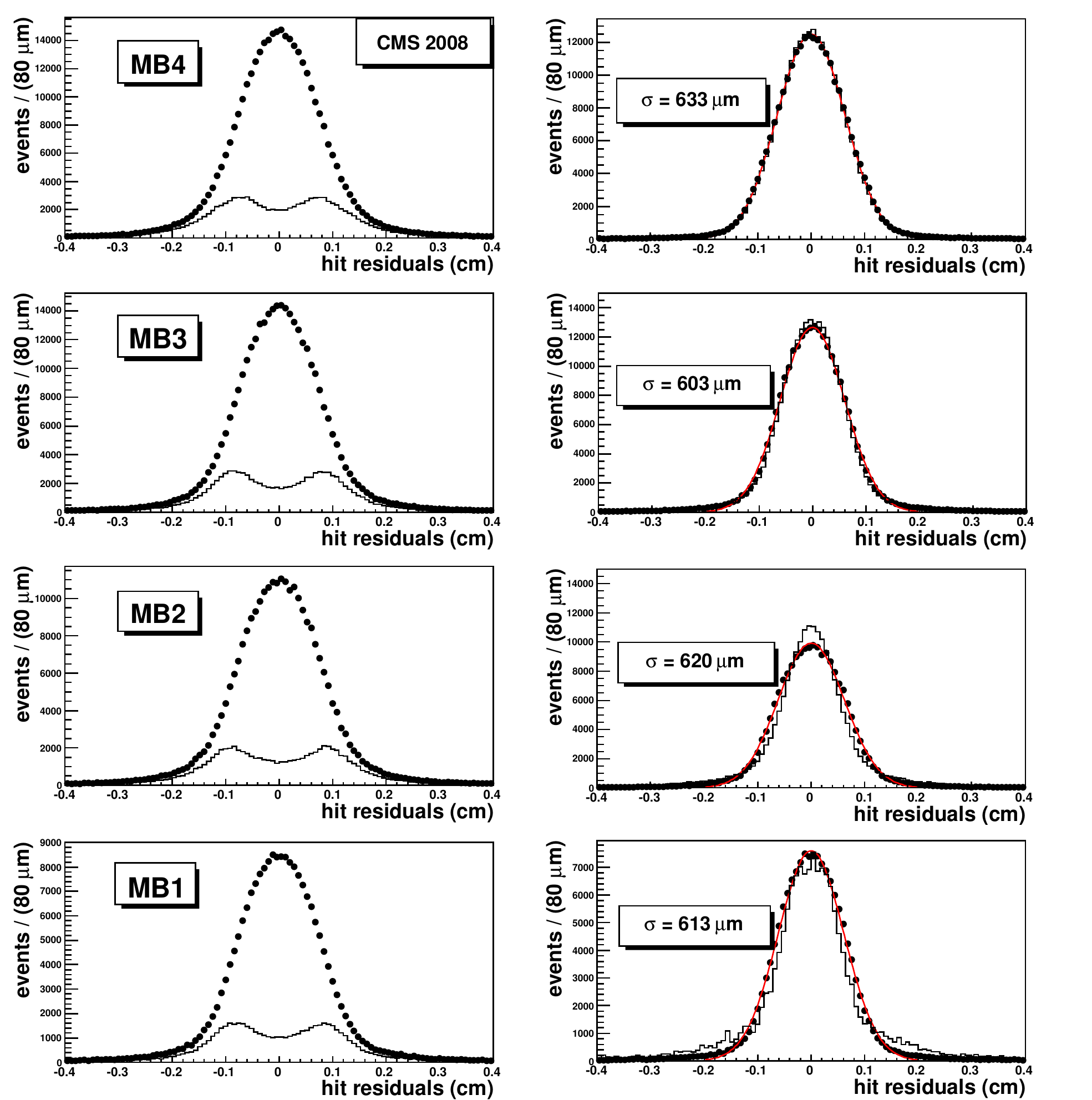}}

  \caption{Hit residuals in DT muon chambers of YB$0$, sector 4, at the first stage of the hit
  reconstruction. Left column plots:
   all events; the full line histograms show the hit residuals for
   the events with bunch crossing identification in the chamber different from the most frequent one.
   Right column: events with the most frequent bunch crossing identification; real data: points,
   simulated data: full histogram. The curves show the result of a fit to the data using a Gaussian function.
   The fitted RMS values are listed. }
  \label{HitResNot0}
 \end{center}
\end{figure}

A measure of hit resolution is provided by the residuals
of the hit position with respect to the predicted position in the layer obtained from the segments,
reconstructed excluding the hit under study from the fit. The distribution of the residuals
in the $r$-$\phi$ SL´s with respect to the position obtained from the segment extrapolation
is shown in Fig.~\ref{HitResNot0},
for the first stage of the hit reconstruction. The data are shown for the four stations of sector 4 in the central wheel of the barrel detector.
Only segments with more than 6 hits used in the fit were considered.
The full line histograms shown in the left plots in the figure correspond to the hit residual
distributions from ``off-time'' events,
i.e., events triggered with a bunch crossing identification provided by the local trigger of the chamber differing
by $\pm 1$ (in $25$~ns units) from the one occurring more frequently.
As expected, for this population of events the spread of the residuals is significantly larger, since the subtracted time pedestal computed by the
calibration procedure is shifted on average by $\pm 25$~ns with respect to the muon arrival time. The double
peak structure for these events reflects the staggering of the DT cells between consecutive layers:
hits occurring on the half-cell volume on the left side of the anode wire have a bias opposite with respect to hits
occurring in the half-cell volume on the right side.

In the right plots of Fig.~\ref{HitResNot0} the distribution of the residuals is shown
both for real and simulated data for ``in-time'' events, i.e., for events triggered with the most frequent bunch crossing identification
in the chamber.
A single Gaussian fit to the residual distributions, shown by the curve superimposed to the data point, gives
$\sigma_{\mathrm{res}}=620$ $\mu$m. To have an estimation of the hit resolution at this stage, this value must be
corrected for the segment extrapolation error, which at this reconstruction stage
is on average $\sigma_{\mathrm{extrap}}=320$ $\mu$m (slightly dependent on the
layer position of the hit under test). The observed single hit resolution is thus:

 \hspace{1cm} (1)\hspace{2cm}  $\sigma_{\mathrm{hit}}= [\sigma_{\mathrm{res}}^2 - \sigma_{\mathrm{extrap}}^2]^{1/2}=$ $530$ $\mu$m.

The pedestal-subtracted time recorded by the TDC is the sum of the electron drift time (ranging from 0 to a maximum of
about 380 ns for muon tracks passing at the DT cell boundary~\cite{MTCCvdrift}), the random
arrival time $t_0$ of the muon in the trigger window and the time of the signal propagation
along the anode wire. This last effect can be taken into account once the segment pattern recognition is
performed in the orthogonal superlayer and
the hit position along the wire is determined. The expected hit resolution is then:

 \hspace{1cm} (2) \hspace{2cm} $\sigma_{\mathrm{hit}} = [\sigma_{\mathrm{cell}}^2+\sigma_{t_0}^2+\sigma_{\mathrm{prop}}^2]^{1/2}=$ $470$ $\mu$m

roughly consistent with the observed value. In the expression
above, $\sigma_{\mathrm{cell}}=$ $200$~$\mu$m is the intrinsic
position resolution of the DT cell as measured with muon test beam~\cite{DTtest1} and
$\sigma_{t_0}=(25$~ns~$/\sqrt{12})\cdot v_{\mathrm{drift}}=$ $390$~$\mu$m is the contribution due to the
uncertainty of the muon arrival time for an average electron drift
velocity $v_{\mathrm{drift}}=$ $54$~$\mu$m/ns~\cite{DTtest1}. Finally
$\sigma_{\mathrm{prop}}=v_{\mathrm{drift}} \cdot \sigma_t=160$~$\mu$m is the uncertainty due to the
signal propagation along the anode wire, where
$\sigma_t=(l/\sqrt{12}) / v_{\mathrm{prop}}$, $v_{\mathrm{prop}}=0.244$ m/ns is the signal propagation
velocity~\cite{MB3test} and
$l = 2.5$~m is the anode wire length. The corrections with respect to the ideal detector geometry
for the layer misalignments inside the chambers~\cite{CFTalisoftw} have been included in the
reconstruction. The contribution to the observed hit resolution from the remaining uncertainty
(of the order of $30$-$40$~$\mu$m) on this corrections is negligible.

\begin{figure}[htbp]
 \begin{center}
  \resizebox{8cm}{!}{\includegraphics{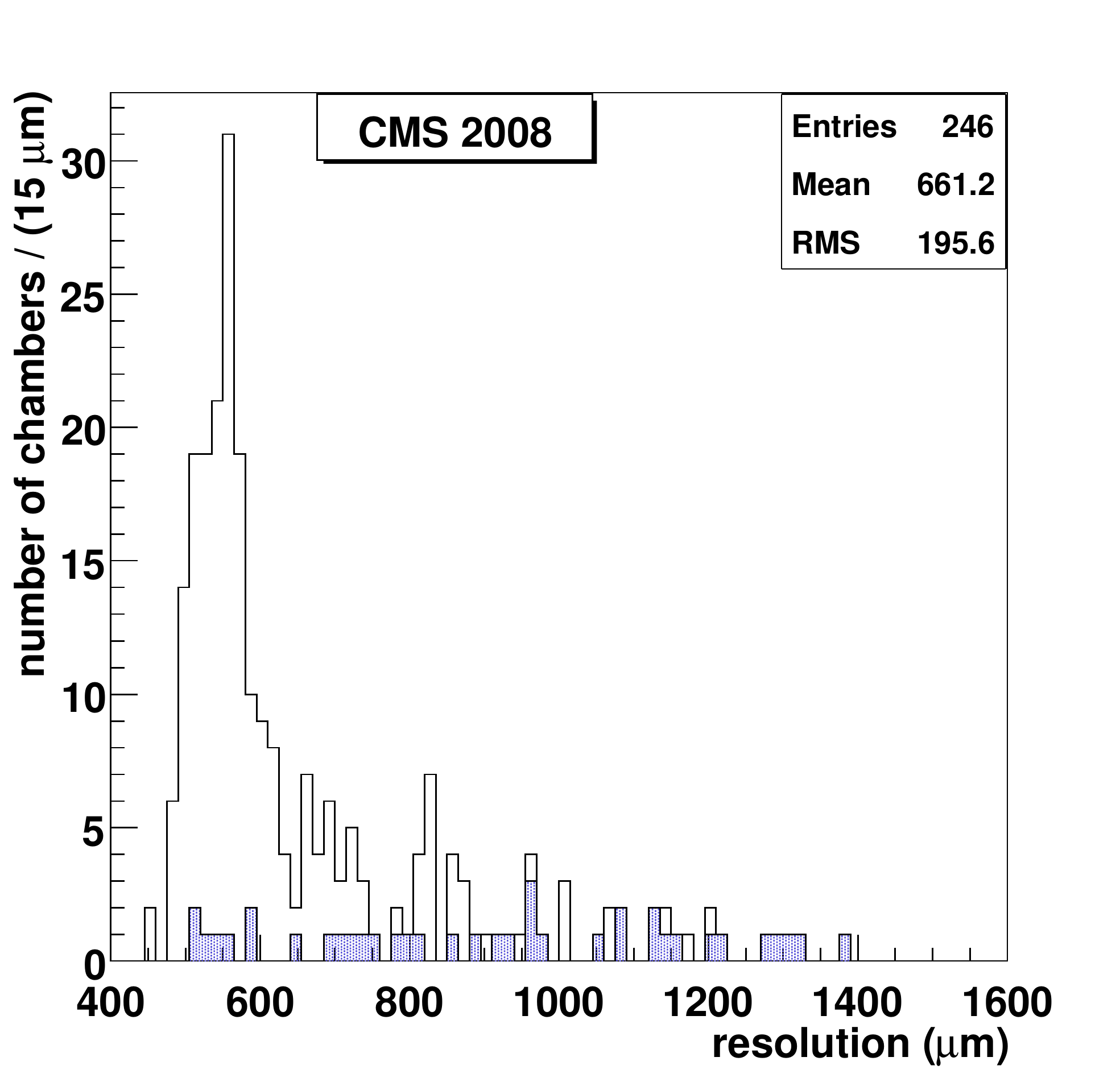}}
  \caption{Distribution of the hit resolution computed using Eq. (1) from the RMS values
  of the Gaussian function fitted to the reconstructed hit residuals in all DT chambers, obtained
  at the first stage of the local reconstruction.
  The dark entries are from chambers in the vertical sectors. Four chambers are not included in the plot due to powering problems.}
  \label{HitResAllchNot0}
 \end{center}
\end{figure}

The distribution of the hit resolution, obtained using Eq. (1) from
the RMS values of the Gaussian function fit to the hit residuals, is shown in
Fig.~\ref{HitResAllchNot0}. The average value of the
distribution obtained for 246 chambers is $660$ $\mu$m with an RMS of about $200$ $\mu$m.
In addition to the two chambers completely switched off, there
were two chambers in sector 8 of YB$1$ and YB$-1$ respectively having the innermost $r$-$\phi$ SL
switched off (cfr. Fig.~\ref{fig:RecHitEffSummary}), for which the hit resolution study
was not performed. It is worth noting that
the tail in the distribution comes from the chambers in the most inclined sectors with respect to the
horizontal direction. In particular, the worst
performance is obtained in the chambers of the vertical sectors 1 and~7 (corresponding to the shaded entries shown in the histogram), where the
average direction of the triggered cosmic muons with respect to the chamber normal axis is
larger than $50$ degrees. In this condition, which is very far
from the one expected for prompt muons originating in pp collisions at the LHC, the
$t_{\mathrm{trig}}$ determination has larger uncertainties and
the effects due to cell non-linearity become important.

\begin{figure}[htbp]
 \begin{center}
  \resizebox{17cm}{!}{\includegraphics{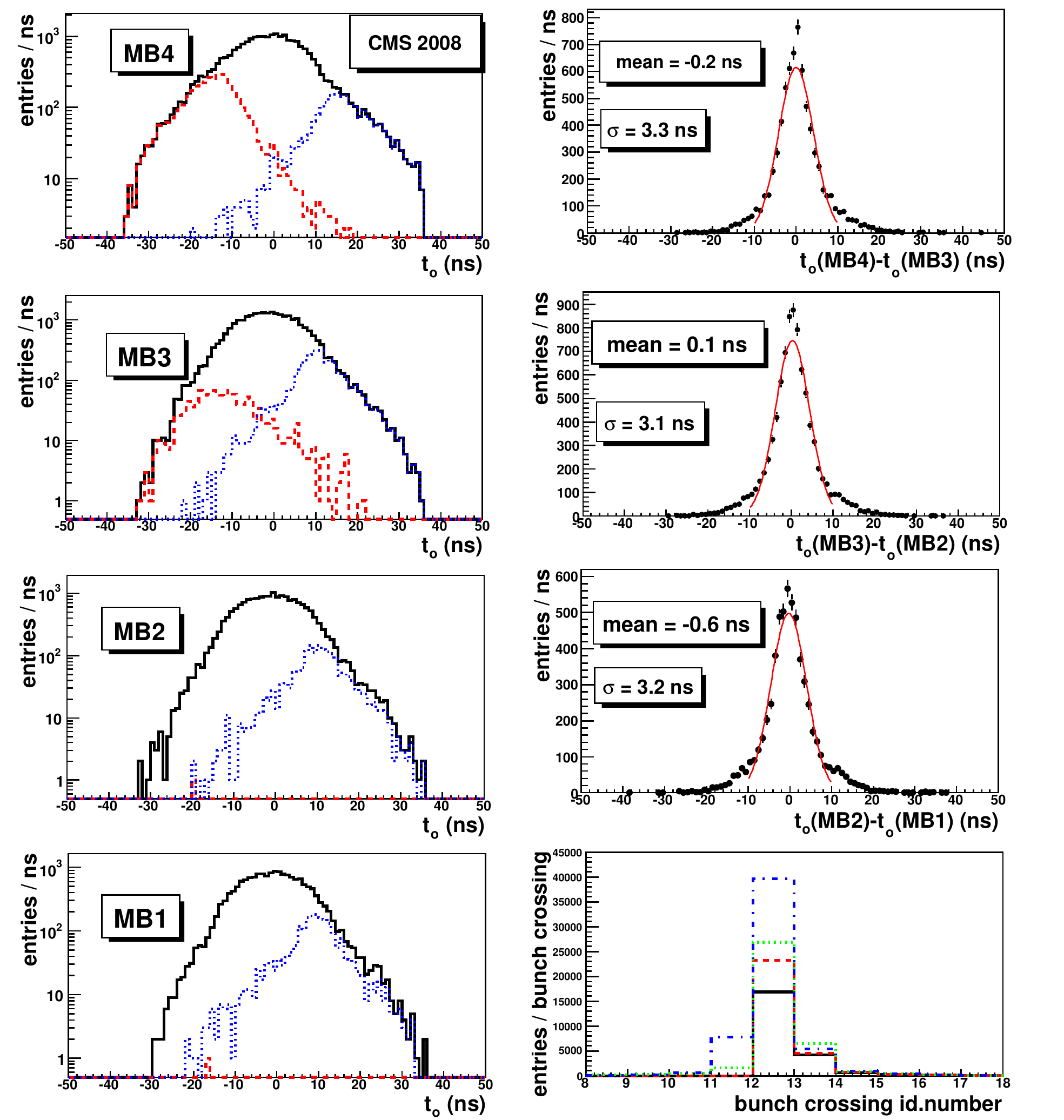}}
  \caption{Left column: distributions of the fitted arrival times of the muon in the chambers of sector 4
  in YB$-1$. The full line histograms refer to all events triggered by the local trigger. The dotted (dashed) line
  histograms refer to events with bunch crossing identification = +1 (-1)
  with respect to the most frequent bunch crossing (12)
   provided by the local trigger in each chamber~\cite{DTtrigg}.
   Three upper right plots: distributions of the difference of the $t_0$ values between two consecutive stations.
   The curves show the result of a Gaussian fit over the range [-10,+10]~ns. The
   fit results are given to provide a rough measure of the mean and RMS of the core of the distribution.
   Bottom right plot:
   distributions of the bunch crossing identification in the four chambers of the sector (full line histogram: MB1; dashed line: MB2;
   dotted line: MB3; dashed-dotted line: MB4).}
  \label{t0plots}
 \end{center}
\end{figure}

After the local pattern recognition, the arrival time of the muon, $t_0$, can be treated as a free parameter
in a refit of the segment that determines the final segment position and direction~\cite{MTCCvdrift}.
Typical distributions of the fitted muon arrival time in the chambers of sector 4 are shown
in Fig.~\ref{t0plots}, for all events triggered by the local trigger, and separately for bunch crossings
differing from the most common by one.
The local trigger assigns the candidate track to a given bunch crossing time window,
defined with 25 ns granularity. The distributions of the bunch crossing identification number in all the
chambers of the sector are also shown in Fig.~\ref{t0plots}. Although the number is arbitrary, it is
evident that the tails are dominated by events triggered at the bunch crossing differing by $\pm 1$ from
the most commonly identified crossing of 12.
The differences between the distributions of the bunch crossing identification shown
for different chambers in the lowest right plot are due to the imperfect
fine tuning of the synchronization of the local trigger devices of the chambers~\cite{FineSync}.
In this sector, for MB1 and MB2 chambers, the population of events with bunch crossing 11 is
practically absent, as a consequence of
the muon time of flight, which enhances the probability to have in these stations a bunch
crossing identification number shifted by $+1$ with respect to the bunch
crossing number assigned by MB3 and MB4.
The differences between the fitted arrival times in consecutive chambers are also shown
in the figure. It must be stressed that the time pedestal calibration procedure mentioned above
is defined by taking into account the
muon time of flight between them. The average values of the distribution of the time differences
between consecutive chambers are thus expected to be zero.

\begin{figure}[htbp]
 \begin{center}
 \resizebox{17cm}{18cm}{\includegraphics{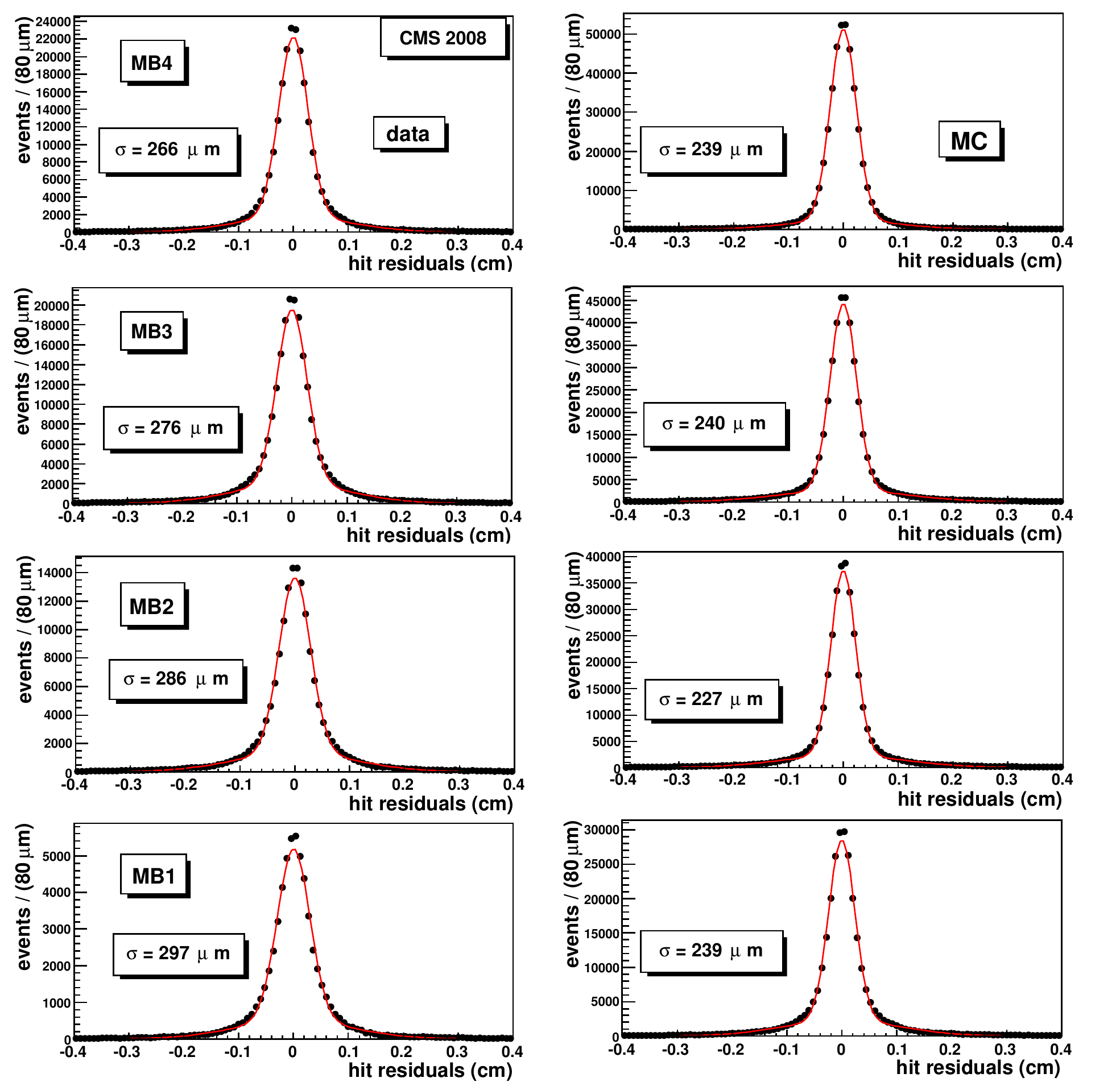}}
  \caption{Hit residuals in DT muon chambers of YB$-2$, sector 4 after $t_0$
  segment refit. Left column: data; right column: simulation.
  The curves show the result of a fit to the data using a double Gaussian function.
   The fitted RMS values of the narrower Gaussian function are listed.}
  \label{HitRes}
 \end{center}
\end{figure}

\begin{figure}[htbp]
 \begin{center}
 \resizebox{17cm}{18cm}{\includegraphics{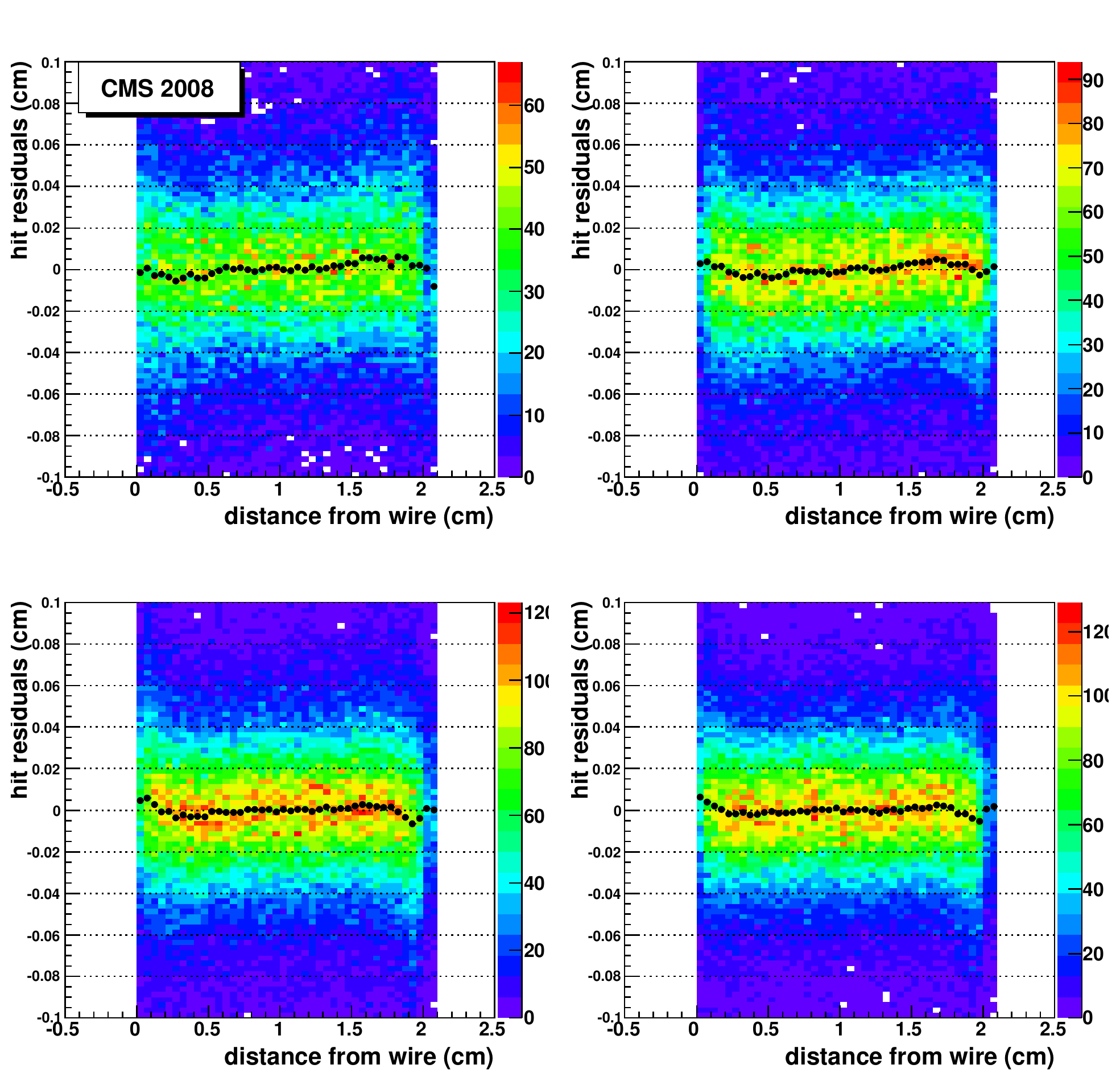}}
 \caption{ Plot of residuals vs hit position in a DT cell, for the chambers of YB$-2$, sector 4; the plot profile is shown by the points.
 Top plots: MB1 (left) and MB2 (right). Bottom plots: MB3 (left) and MB4 (right).}
  \label{HitResVSx}
 \end{center}
\end{figure}

The distribution of the hit residuals after the $t_0$ refit is shown in Fig.~\ref{HitRes} for sector 4
of the external wheel YB$-2$.
In this wheel (as well as in wheel YB$2$), the residual magnetic field in the chambers volume has the
largest variation along the chamber´s length,
reaching the highest values (up to $0.8$~T for the radial component in the MB1 stations~\cite{MUtdr}).
This variation does not affect significantly the average hit resolution observed in the chamber,
once the corresponding average change of the effective electron drift velocity
(about $2\%$ for MB1 chambers~\cite{DTcalibCRAFT}) is taken into account in the reconstruction.
As for the distributions shown in Fig.~\ref{HitResNot0}, the residuals
are computed with respect to the extrapolated position from the segment, obtained excluding
the hit under study. The residuals are shown for all the triggered events.
Plots of the hit residuals vs. the distance to the anode wire in the DT cells are shown in
Fig.~\ref{HitResVSx}, displaying the good uniformity of the cell behaviour in the whole
drift volume. Moreover, the approximate straight line behaviour of the
mean value of the residual distribution in each bin
demonstrates that non-linear effects are smaller than $100$~$\mu$m.
This is in agreement with accurate studies performed on dedicated test beam data, that show
deviations from linearity not larger than $60$~$\mu$m~\cite{DTtest1}.
Although the distributions of hit residuals have width significantly narrower than the
corresponding distributions obtained before the $t_0$ fit, they still have rather large tails.
These are due to displaced hits from $\delta$-rays, originally included in the segment by the
pattern recognition algorithm. It is worth remembering here that the algorithm was run with a
loose criterion to include a hit in the segment, in order to cope with the initial uncertainty on the
hit position dominated by the $t_0$ jitter. The distributions of hit residuals were fitted with
a sum of two Gaussian functions, constrained to have the same mean values.
As seen in Fig.~\ref{HitRes}, the narrower Gaussian gives $\sigma \approx  280$~$\mu$m, accounting for about $80\%$ of the 
total population, while the wider Gaussian has $\sigma \approx 1$~mm.

\begin{figure}[htbp]
 \begin{center}
 \resizebox{8cm}{!}{\includegraphics{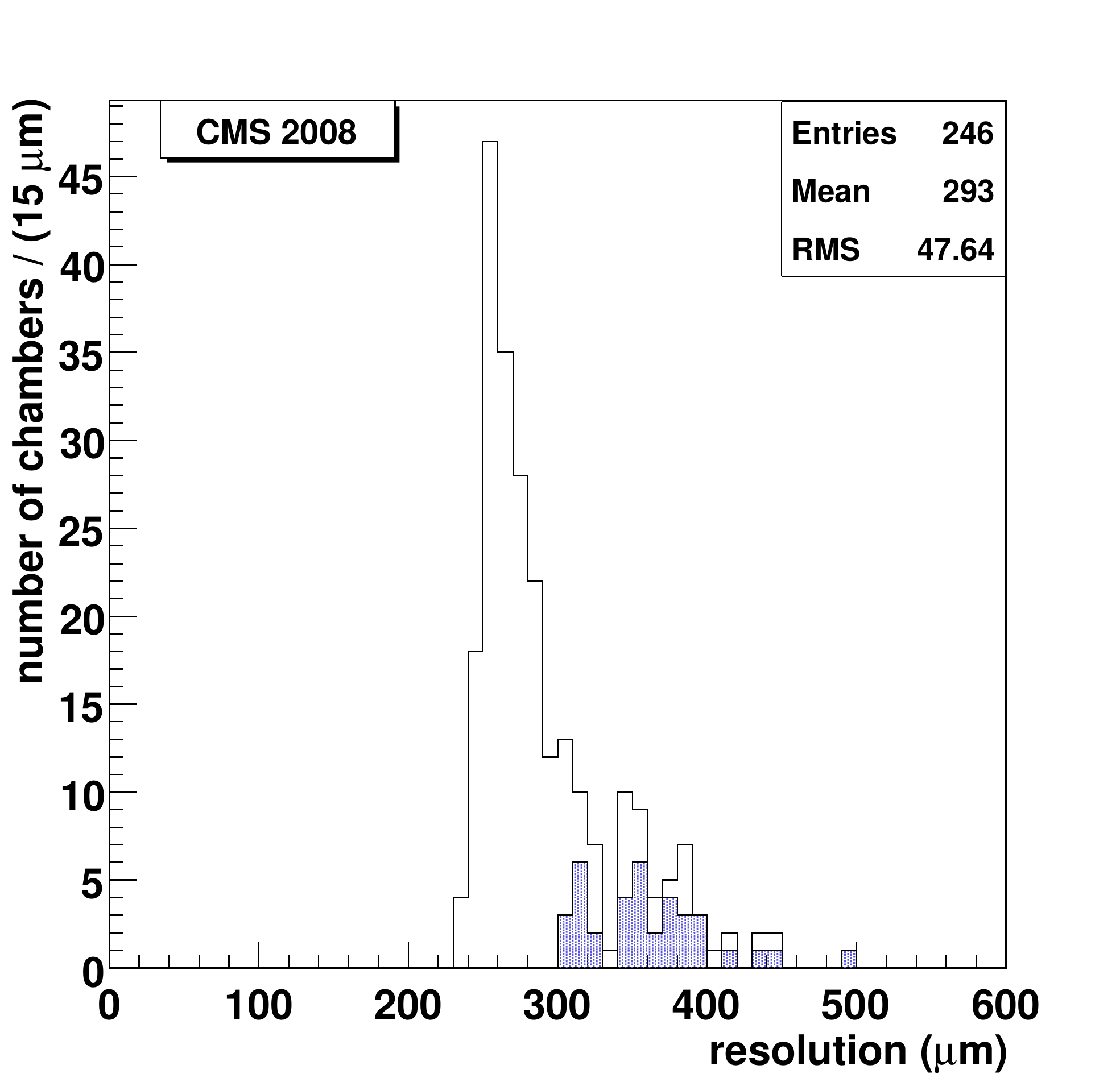}}
  \caption{Distribution of the RMS values of the narrower Gaussian curve fitted to the reconstructed hit residuals in
  all DT chambers, after $t_0$ segment refit.
  The plotted values have been corrected for the track extrapolation error.
  The dark entries are from chambers in the vertical sectors.}
  \label{AllFitRes}
 \end{center}
\end{figure}

The distribution of the hit resolution, computed using Eq. (1) from the RMS values of the
narrower Gaussian function fitted to the reconstructed hit residuals
in all the DT chambers, is shown in Fig.~\ref{AllFitRes}. The value of the extrapolation error
used in Eq. (1) is $\sigma_{\mathrm{extrap}}=140$~$\mu$m.
For most of the chambers, the resolution is approximately $260$~$\mu$m.
Again, the tail at large values comes from chambers in the sectors most inclined with respect to
the horizontal direction. The shaded entries in the histogram are from vertical chambers.

\subsection{Hit Reconstruction Efficiency}

\begin{figure}[htbp]
 \begin{center}
\resizebox{12cm}{!}{\includegraphics{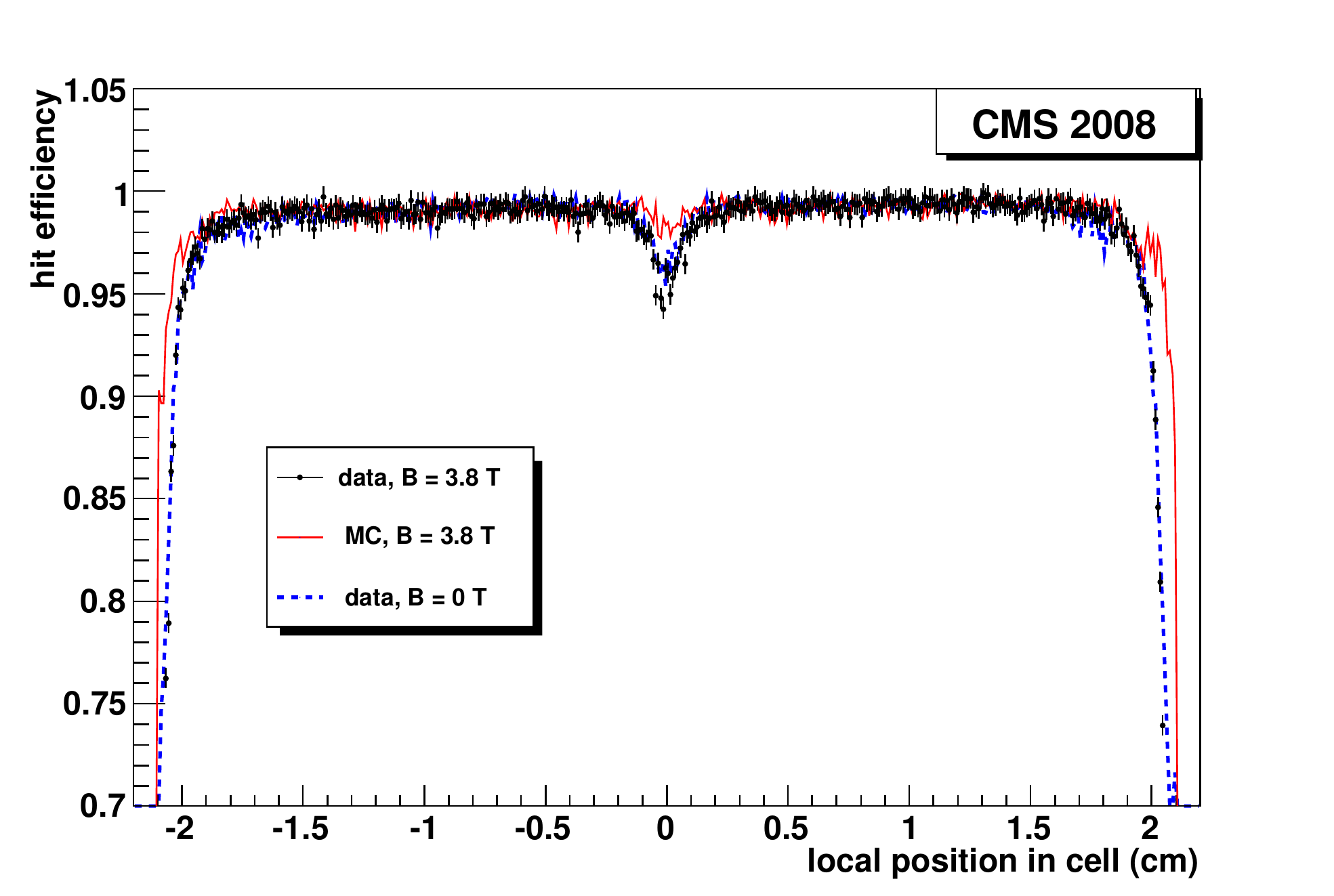}}
  \caption{ Efficiency to have reconstructed a hit in a cell crossed by a cosmic muon, as a function of the
  predicted muon position in the cell,
  for the MB1 stations. The $x=0$ position corresponds to the location of the anode wire in the cell.}
  \label{fig:RecHitEff}
 \end{center}
\end{figure}

The hit reconstruction efficiency is measured by looking for hits in a given layer after extrapolating
the local segment fit to that layer. The extrapolation is done with hits on the segment after excluding
in the reconstruction the hits in the layer under consideration.
Figure~\ref{fig:RecHitEff} shows the efficiency as a function of the predicted hit position in the cell for MB1 stations (data from
all the cells from all the chambers of a given type are combined in the plot). The efficiency is greater than $98\%$ over a large
part of the drift volume. Similar behaviour is observed for the MB2--4 stations.
The observed small inefficiency near the anode wire ($x=0$ in the plots) is due to the pedestal subtraction procedure
described in Section 4 and is well reproduced by the simulation. However, near
the cell boundaries the efficiency is overestimated by the simulation in the last millimeter of the cell volume (corresponding to $5\%$ of the total sensitive
volume). No significant difference between the data at $B=0$~T and $B=3.8$~T is observed.
The noise effect is negligible in this plot because the number of noisy cells having an occupancy
larger than $1\%$ in the recorded data amounts to less than $0.1\%$ of the total number
of DT cells. A detailed study of noise rates in the DT system can be found in Ref.~\cite{DTcalib}.

\begin{figure}
 \begin{center}
%
\resizebox{15cm}{!}{\includegraphics{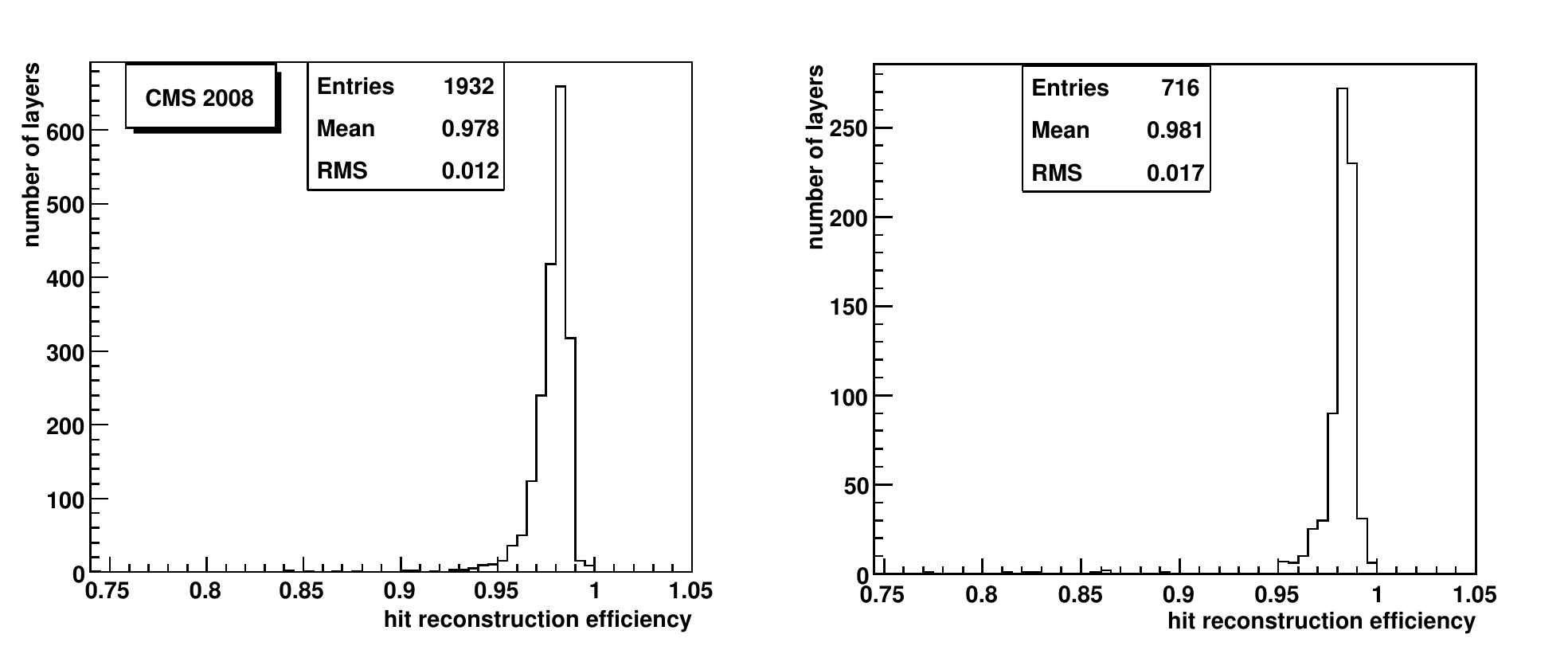}}
  \caption{ Average of the reconstructed hit efficiency in the layers of the Muon Barrel DT chambers. Left:
  $r$-$\phi$ superlayers; right: $r$-$z$ superlayers.}
  \label{fig:RecHitEffSummary}
 \end{center}
 \end{figure}

Figure~\ref{fig:RecHitEffSummary} summarizes the results for the hit efficiency in all the layers
of the DT chambers, averaged over all the cells of the considered layer. The efficiency is higher than $95\%$  almost everywhere in
the barrel detector, with a small decrease in the vertical sectors.

\section{Reconstructed Track Segments in DT Chambers}

The second stage of the local track reconstruction described in Section 3 provides ``2D'' and ``4D'' track
segments, which are studied in detail in this section.

\subsection{Multiplicity of associated hits and track segment efficiency}

\begin{figure}[htbp]
 \begin{center}
  \resizebox{17cm}{!}{\includegraphics{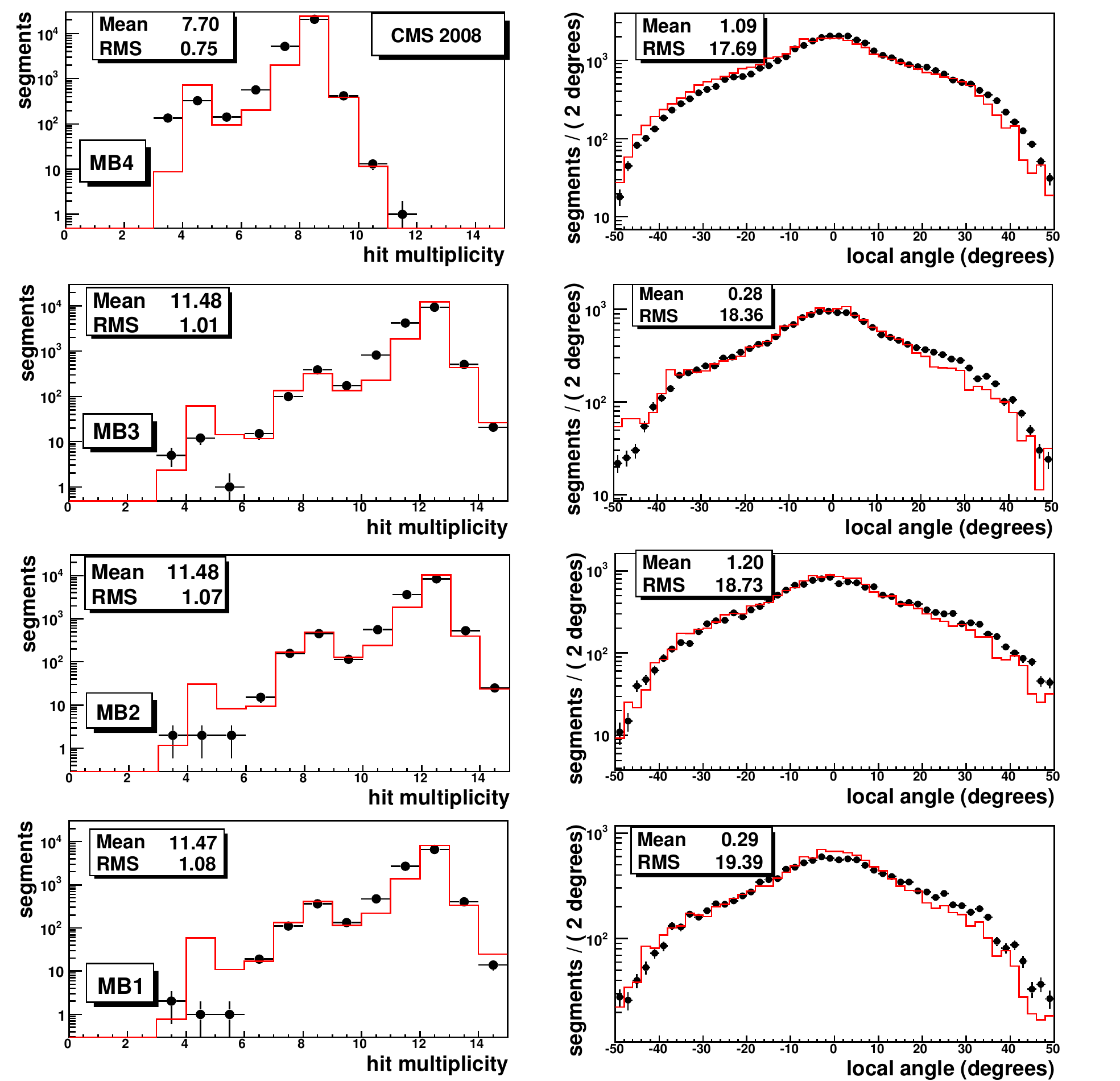}}
  \caption{ Left plots: multiplicity of associated hits in reconstructed 4D segments in YB$+1$, sector 4.
   Right plots: segment direction with respect to the vertical axis. Real data (points) and simulated data (solid line histogram) are shown
   in both sets of plots.}
  \label{fig:HitMult}
 \end{center}
\end{figure}

Reconstructed hits are associated to 2D track segments built independently in the $r$-$\phi$ and
$r$-$z$ planes, as described in Section 3. Collections of 4D track segments are then built considering
all possible combinations of 2D $r$-$\phi$ and $r$-$z$ segments in each chamber.
The distributions of hit multiplicities for all reconstructed 4D track segments
are shown in Fig.~\ref{fig:HitMult} for each DT station in the horizontal sectors of YB$1$ separately.
The distributions are peaked, as expected, at the total number of layers
in the chamber (8 in MB4 and 12 in the other stations), although the Monte Carlo simulation predicts
a slightly larger average multiplicity.
Track segments that have a large incident angle and pass near the boundary between neighbouring drift cells may have more than one associated hit in a
given layer, thus resulting in a hit multiplicity larger than the number of layers in the station.
The distribution  of the segment incident angle with respect to the vertical axis in the bending plane of CMS,
also shown in Fig.~\ref{fig:HitMult}, is well reproduced by the simulation. The observed increase of the
spread around the normal direction when passing from MB4 to MB1, i.e. from the outer to inner stations
(from top to bottom plots in the figure),
is due to the opposite bending effects of the magnetic field in the steel yokes on positive and negative muons.\par
\begin{figure}[htbp]
 \begin{center}
\resizebox{16cm}{12cm}{\includegraphics{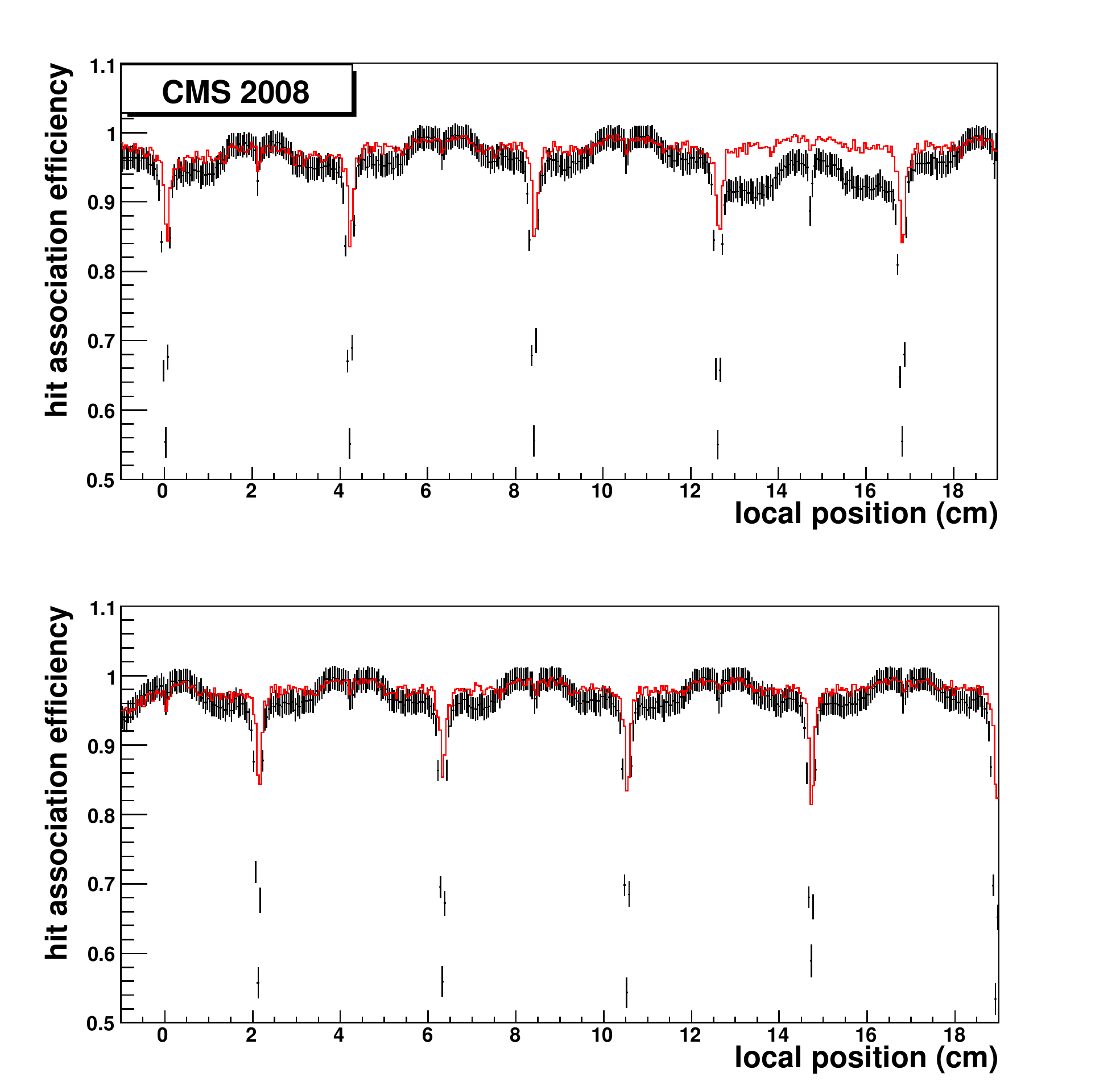}}
  \caption{ Efficiency for hit reconstruction and association to the muon track segment as a function of the predicted muon position
  in layer one (top) and layer two (bottom)
  of one SL in the MB2 station of sector 4 in YB$0$. A region corresponding approximately to four DT cells in each layer is shown.
  Points: data; full line histogram: simulation. Note the suppressed zero of the vertical axis.}
  \label{fig:InEffSingleCell}
 \end{center}
\end{figure}

The difference between data and simulation in the hit multiplicity distributions
is due to the discrepancy in the hit reconstruction efficiency observed near the I-beams separating the DT cells
(see Fig.~\ref{fig:RecHitEff}) and additional small discrepancies,
which sum up independently in the different layers used in the segment reconstruction.
As an example of such small discrepancies, Fig.~\ref{fig:InEffSingleCell} shows the efficiency for hit
reconstruction and association to the muon track, in a region extending approximately over four cells in two
consecutive layers of an $r$-$\phi$ superlayer of the MB2 chamber
in the top sector (sector 4) of YB$0$.
As can be expected, the discrepancy between data and simulation is larger near
the cell boundaries ($0$, $4.2$, $8.4$ ...cm in the first layer shown, staggered by
half a cell between consecutive layers). In addition, a decrease of the efficiency can be due to the
presence of a noisy cell, as is the case for the fourth cell in the upper plot.
A pulse due to noise can indeed mask the hit produced by the muon, which is therefore lost.
Since the number of noisy DT channels is at the level of a few per mille~\cite{DTcalib}, the
overall effect on the multiplicity distributions shown in Fig.~\ref{fig:HitMult} is however negligible.
A discrepancy at a few percent level is also visible
for distances larger than about 1~cm  from the anode wires (located at $2.1$, $6.3$ ... cm in the upper plot),
due to non-linear drift effects.
Finally, the inefficiencies observed very near the anode wires are in general small, especially in
horizontal chambers like the one shown in Fig.~\ref{fig:InEffSingleCell}, for which the time pedestal
determination has a small uncertainty.

\begin{figure}[htbp]
 \begin{center}
\resizebox{10cm}{!}{\includegraphics{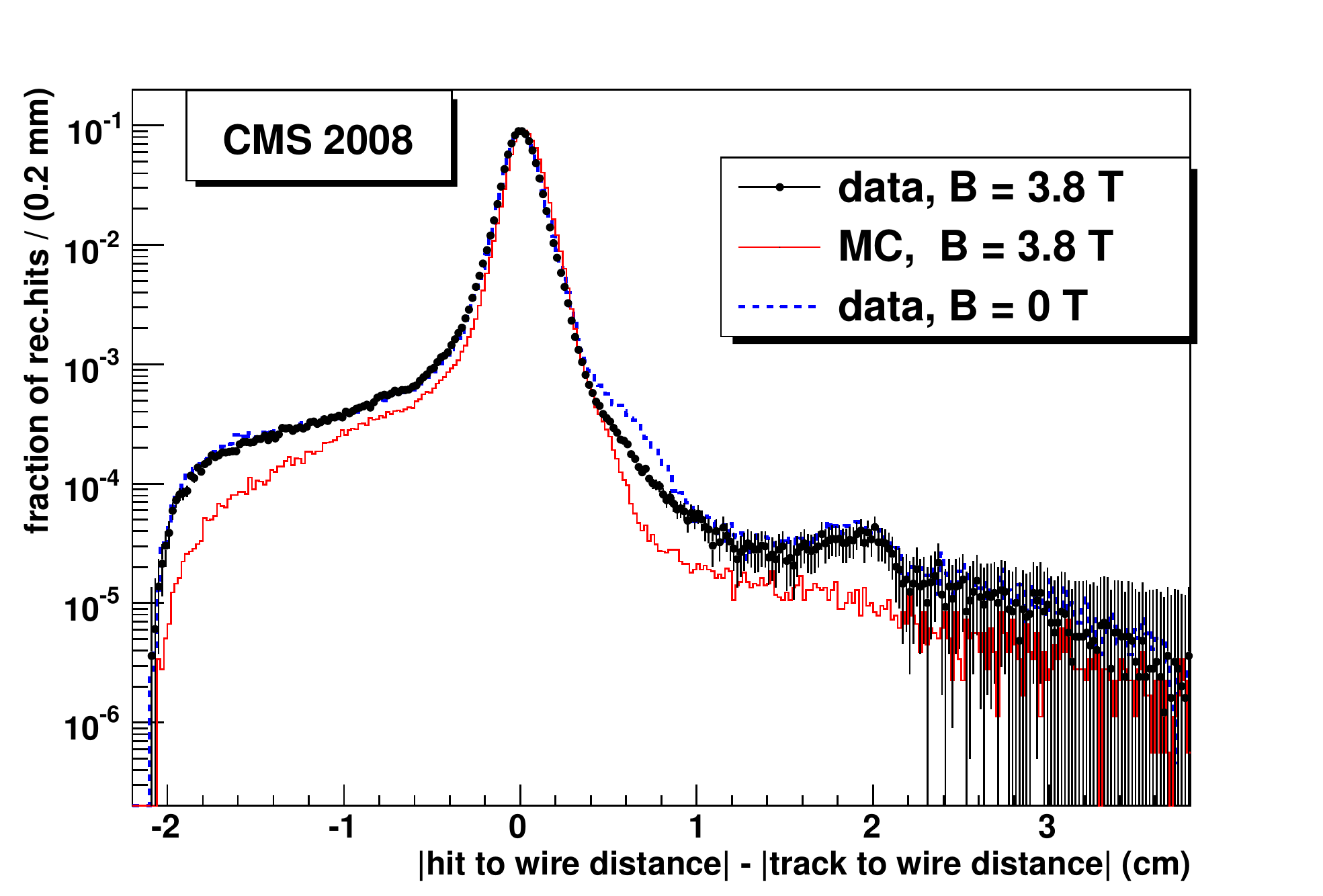}}
  \caption{Distribution of the difference between the distance to the cell anode wire of the first hit recorded in a cell and the distance of
  the extrapolated track position.}
  \label{fig:deltaRays}
 \end{center}
\end{figure}

The efficiency of reconstructed hit association is also affected by the occurrence of $\delta$-ray electrons originating in the gas
volume and/or in the mechanical structure of the chambers. If these electrons pass closer
to the anode wire of the cell than the original muon,
they mask the muon signal if it arrives within the electronics dead time of 150 ns. Figure~\ref{fig:deltaRays}
shows the distribution of the difference between the distance from the cell anode wire of the first hit recorded (independently from its
association to the muon track segment) and the distance of the position
of the track extrapolation. The population at large values of the distance difference is due to the
$\delta$-ray hits that are not associated to the
track segment. The tail at positive values of the difference (extended to values bigger than the half-cell dimension to show the population
from neighbouring cells in the same layer) is due to events with a $\delta$-ray, where the muon hit goes undetected. The data and simulation
distributions show a reasonably good agreement, both in the absolute yield of $\delta$-rays and in the asymmetry
of the distribution, with a slight underestimation of the effect in the simulated data.
The shoulder seen at about $0.8$ cm for $B=0$~T data is due to signals from feed-back electrons 
(see Section 4) extracted from the electrode strip below the anode wire in the cell. This effect 
is almost invisible in the $B= 3.8$~T data, due to the tilt of the electron drift paths which makes 
the detection of these electrons less efficient.
Returning to Fig.~\ref{fig:HitMult}, the difference of about $15\%$ seen in Fig.~\ref{fig:HitMult} between real and
simulated data in the fraction of segments having 12 associated hits (8 in MB4) is understood as
mainly due to an average difference of about $1\%$  in the hit reconstruction and association efficiency, concentrated in the part of the
DT cell farther from the anode wire.

\begin{figure}[htbp]
 \begin{center}
  \resizebox{7cm}{!}{\includegraphics{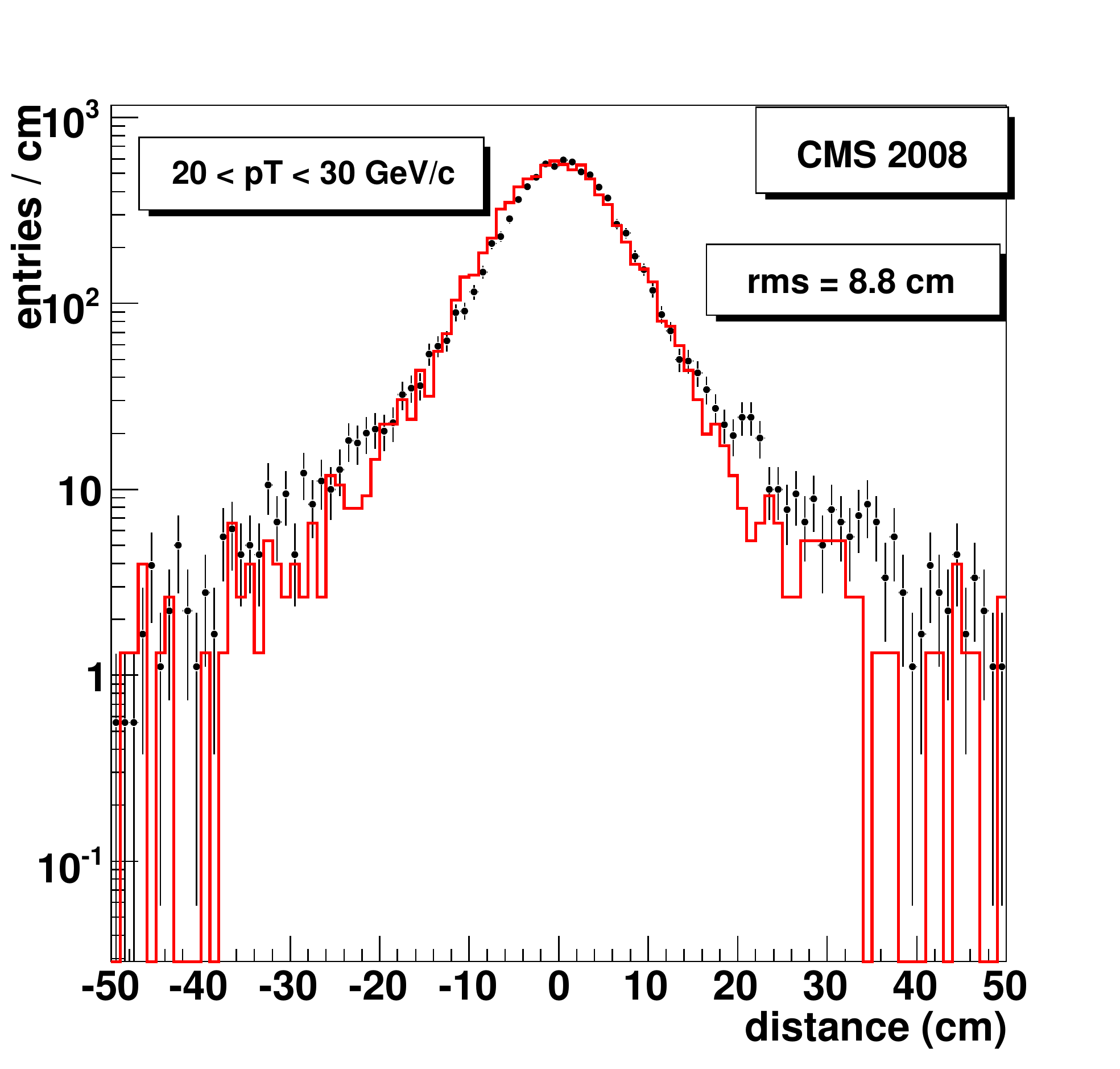}}
  \resizebox{7cm}{!}{\includegraphics{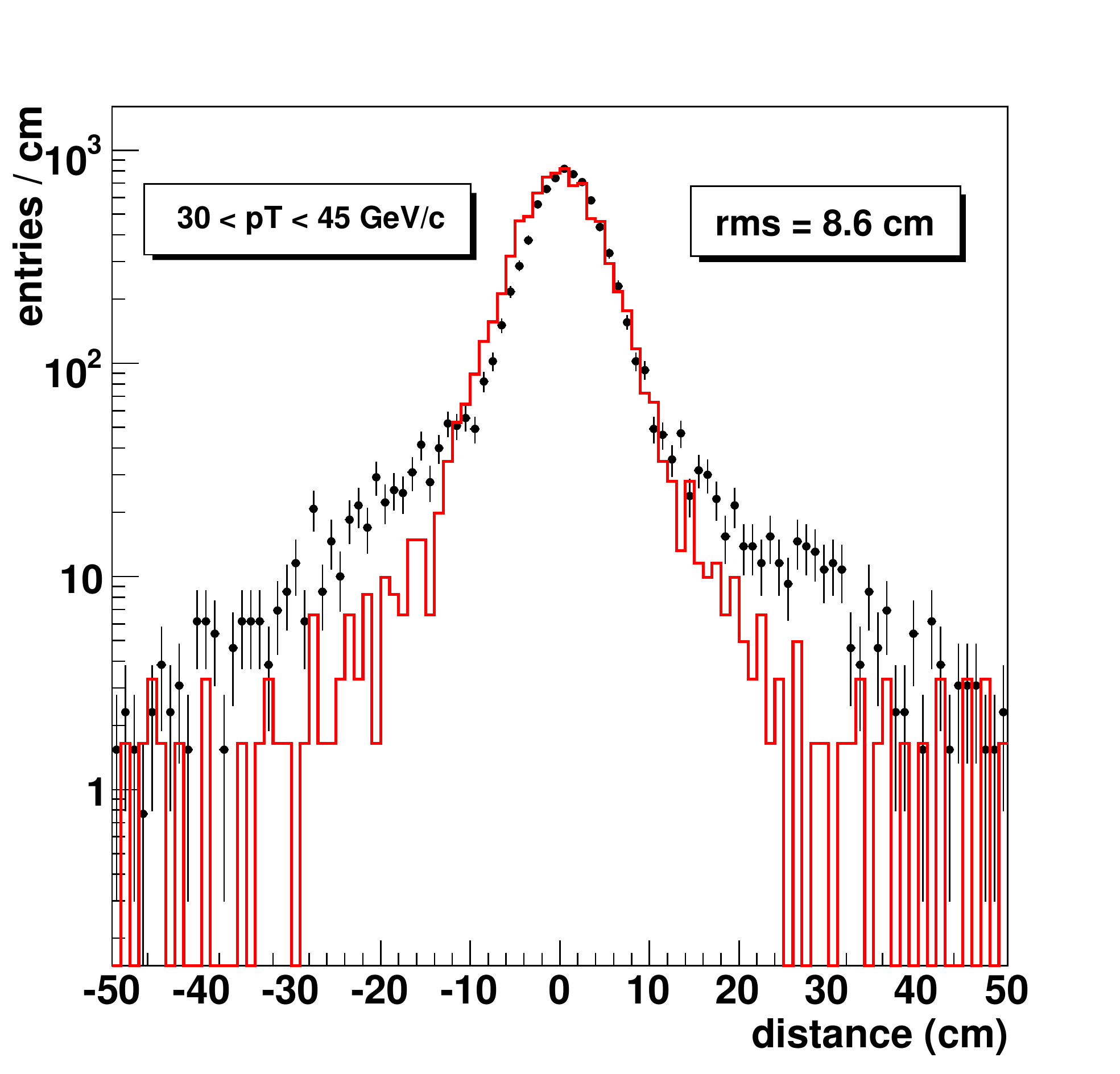}}\\
  \resizebox{7cm}{!}{\includegraphics{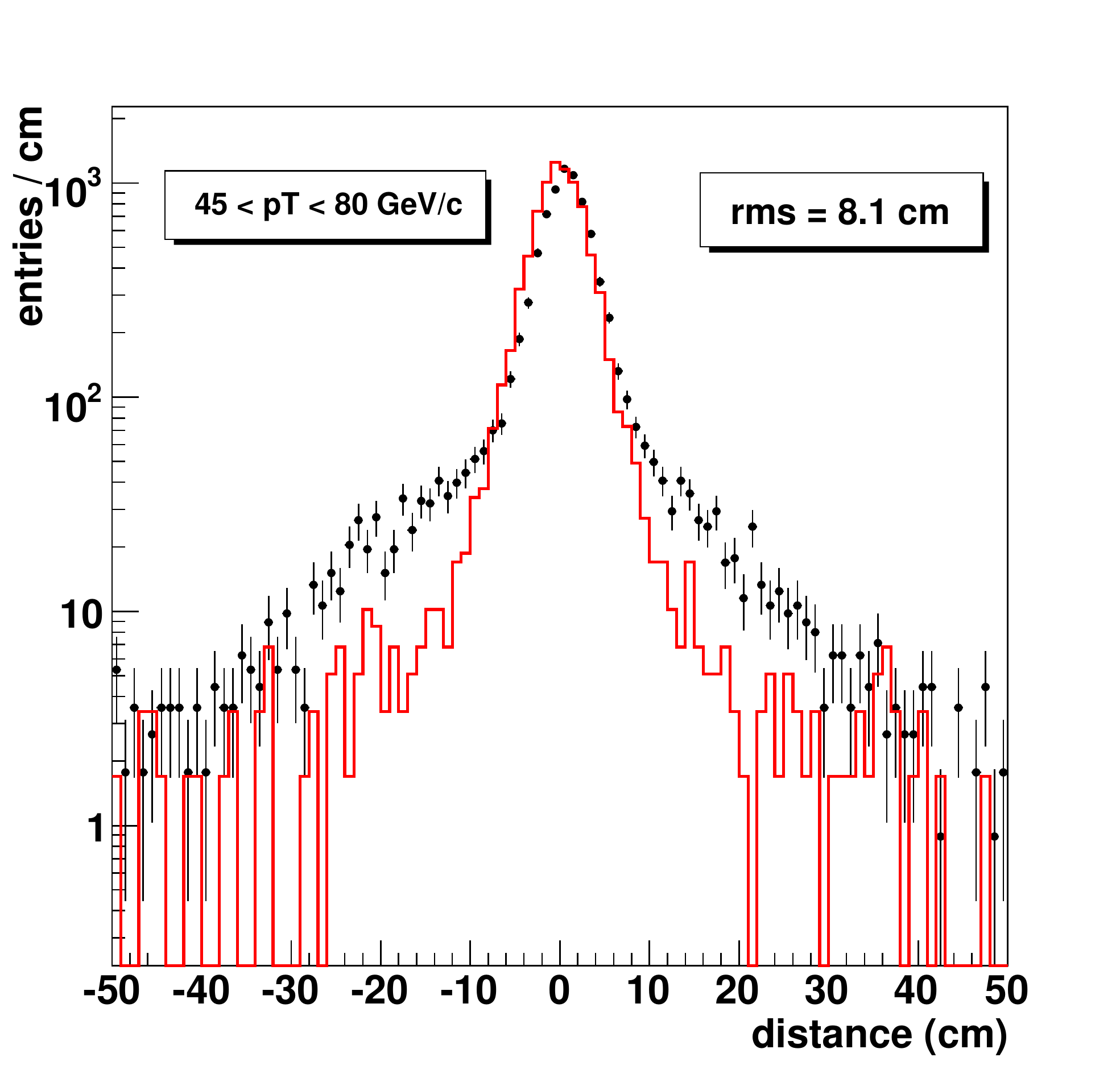}}
  \resizebox{7cm}{!}{\includegraphics{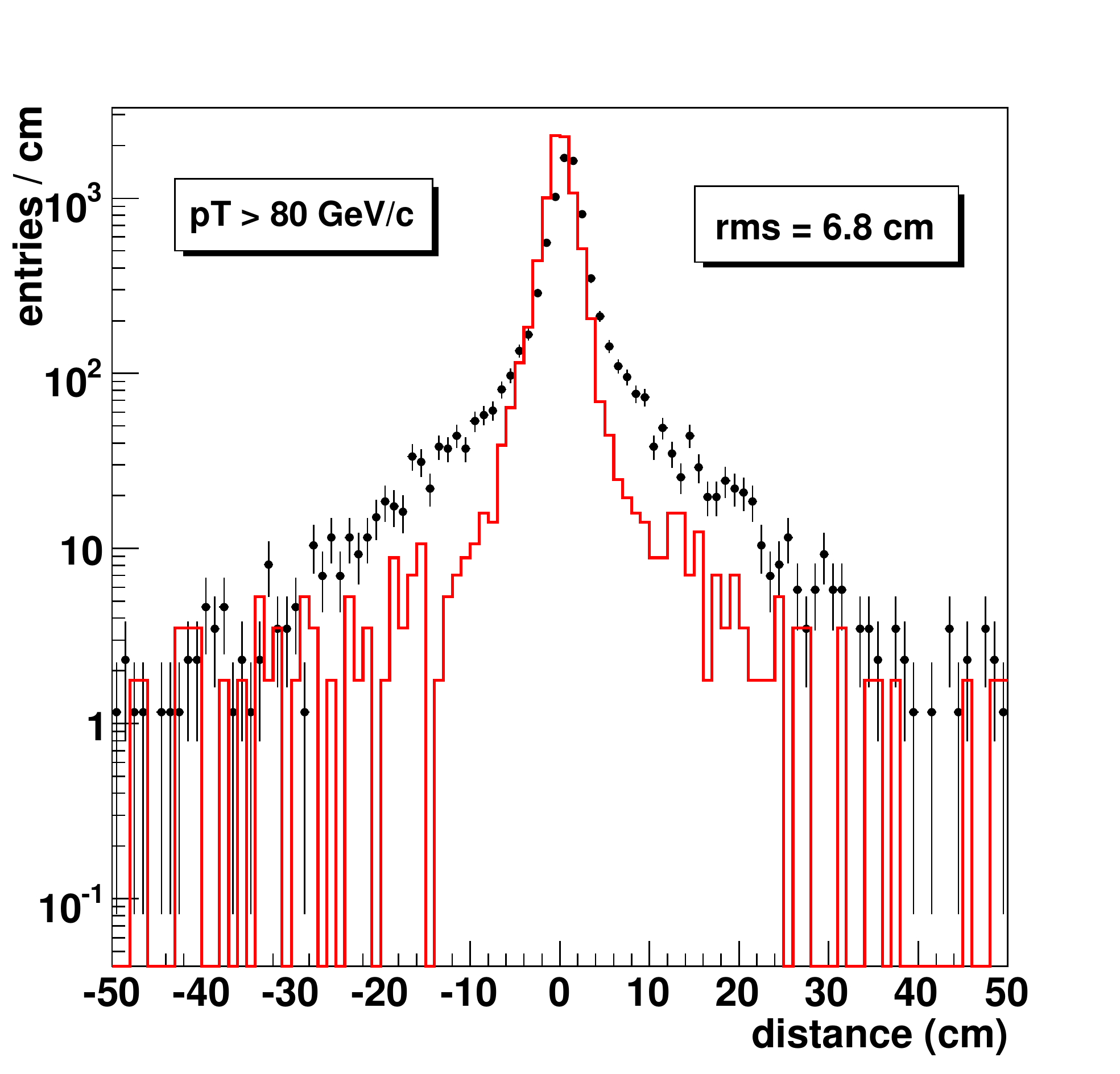}}
  \caption{  Distance between the extrapolated position from the tracker track and the reconstructed
  2D $r$-$\phi$ segment position in MB1, for different $p_T$ bins.
  Dots: real data; full line histograms: simulated data.}
  \label{fig:MB1extrap}
 \end{center}
\end{figure}

\begin{figure}[htbp]
 \begin{center}
  \resizebox{7.5cm}{!}{\includegraphics{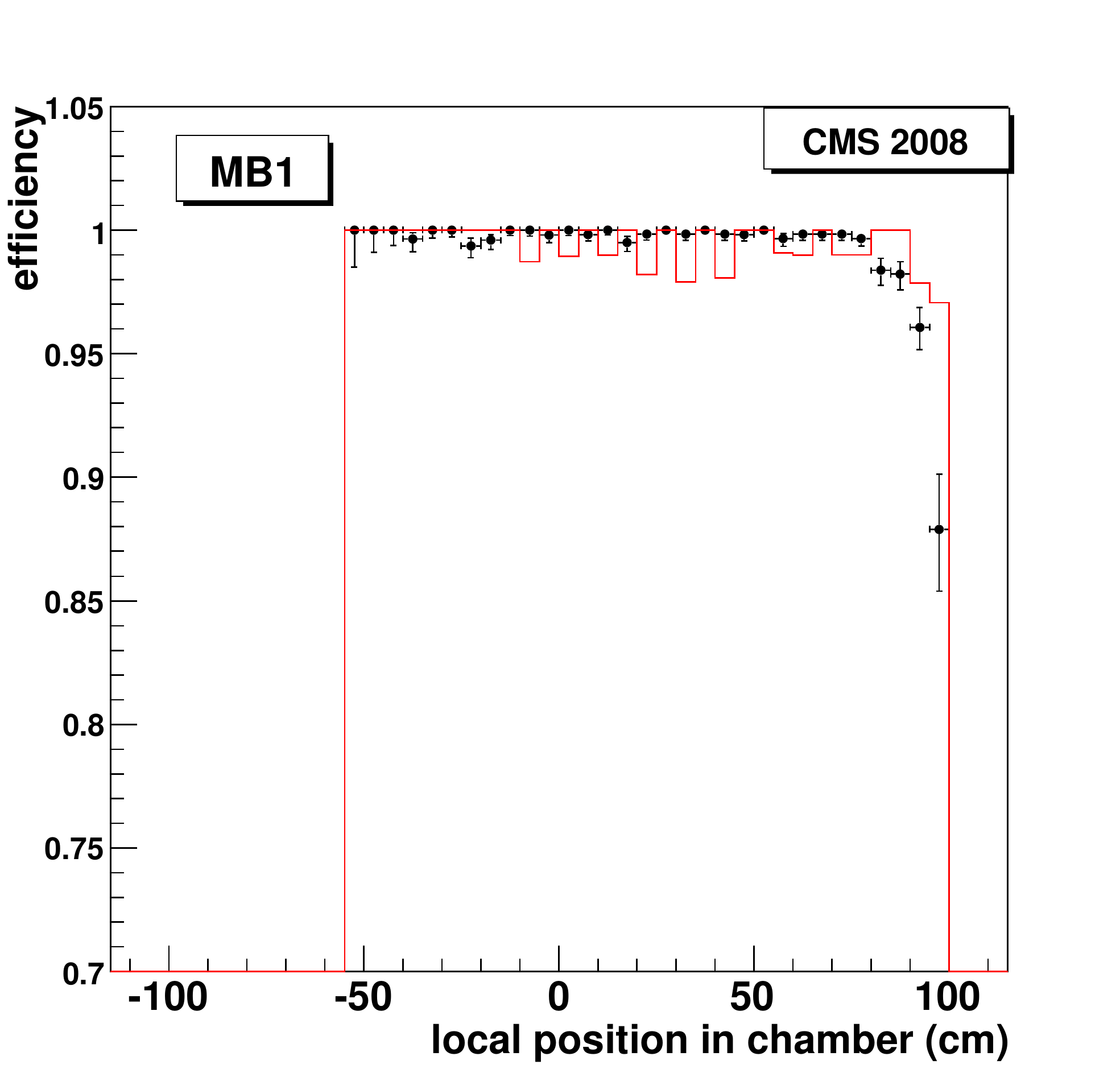}}
  \resizebox{7.5cm}{!}{\includegraphics{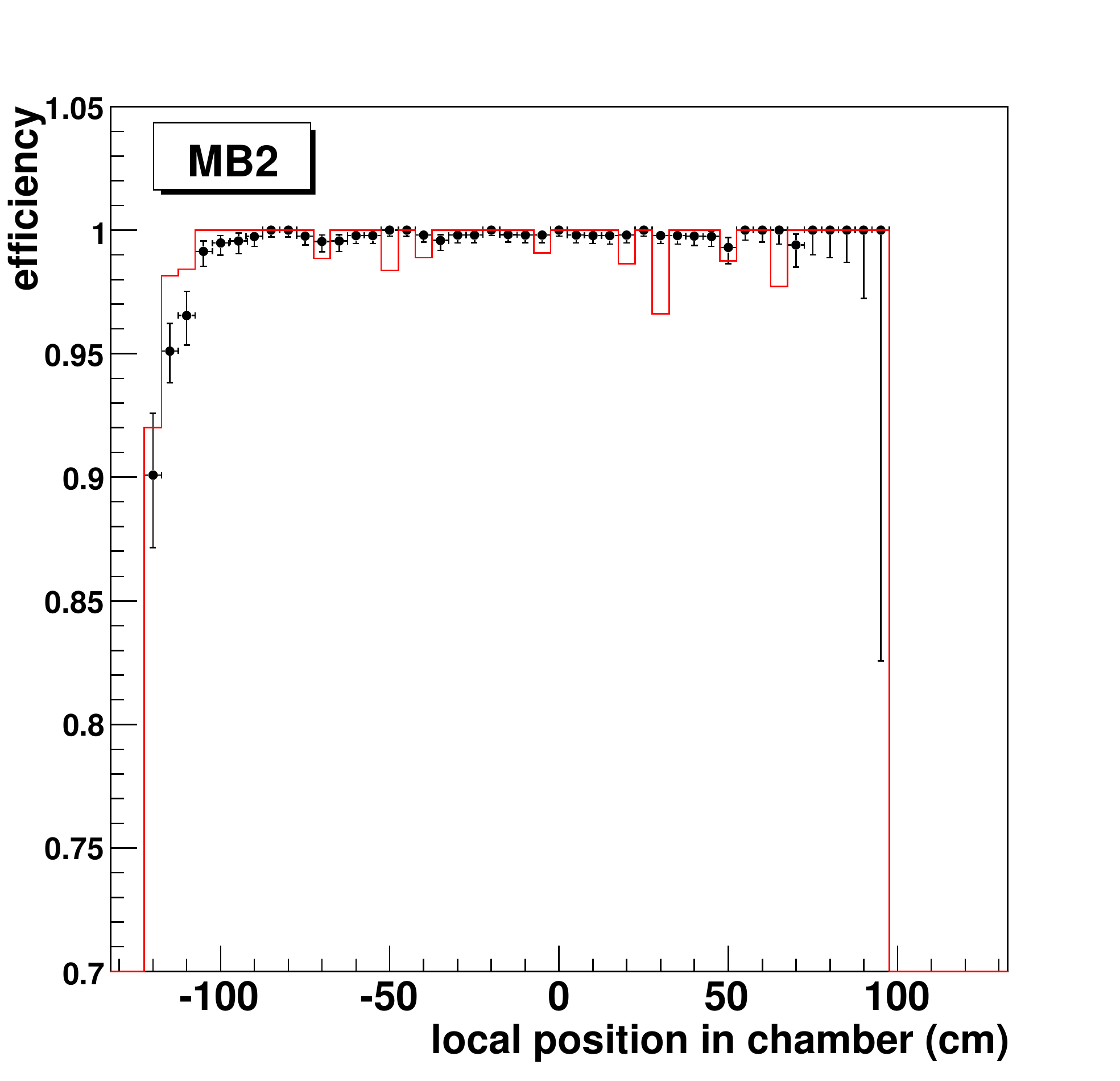}}\\
  \resizebox{7.5cm}{!}{\includegraphics{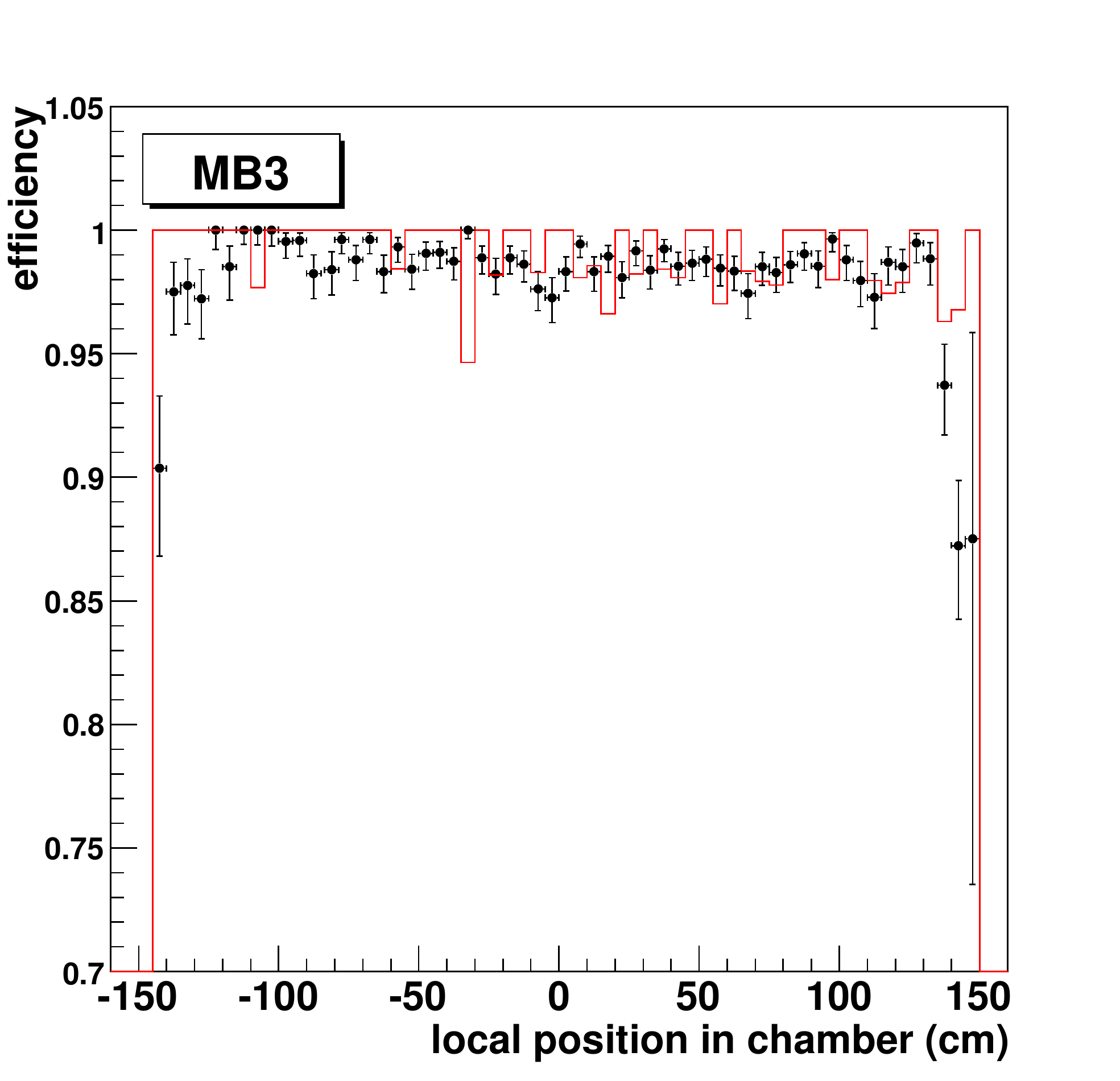}}
  \resizebox{7.5cm}{!}{\includegraphics{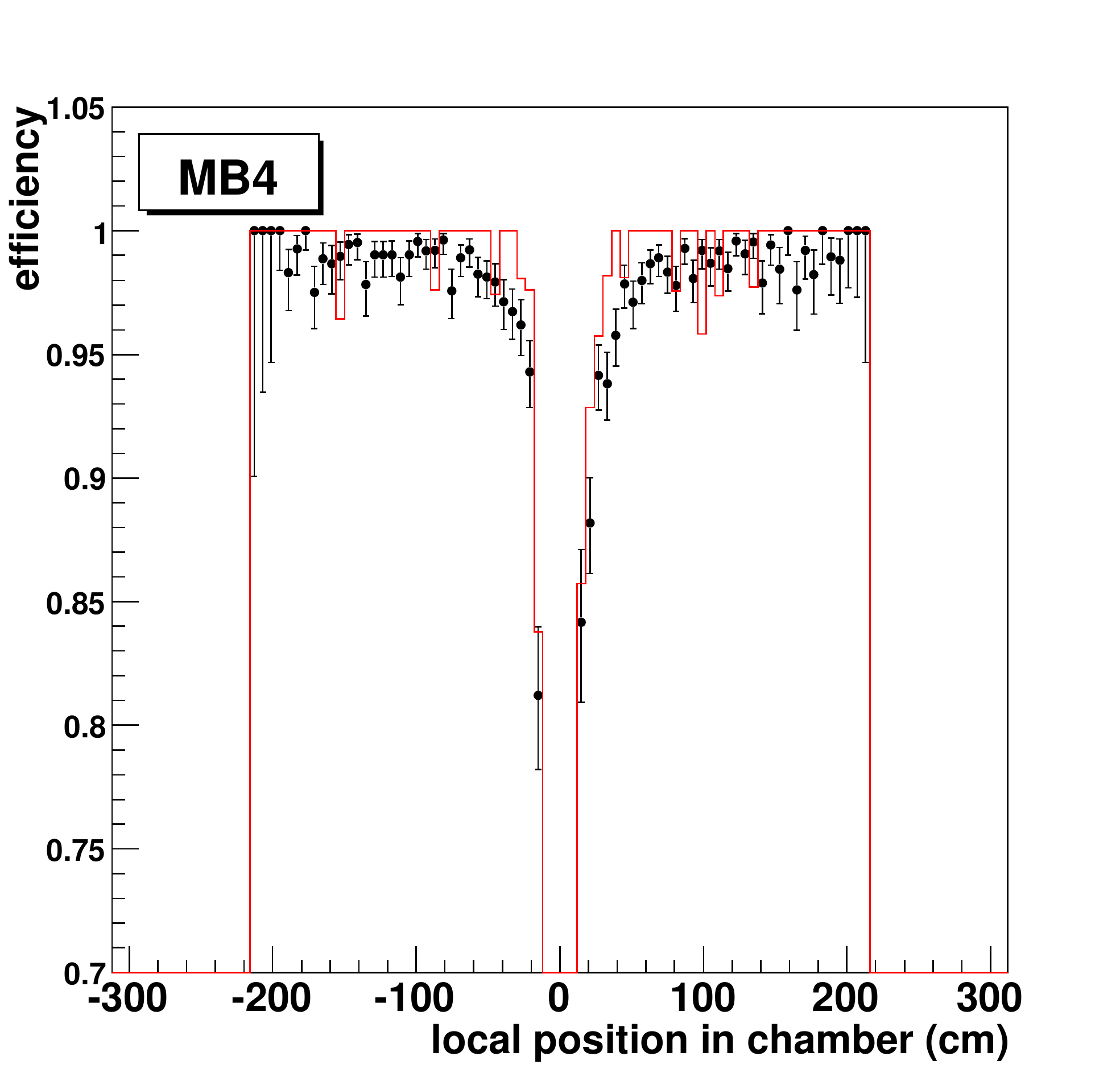}}
  \caption{ 2D $r$-$\phi$ segment efficiency as a function of the local coordinate in the chamber in YB$0$, sector 4.
  Dots: real data; full line histograms: simulated
  data. Note that the MB4 station in this sector is split into two chambers.}
  \label{fig:MB1eff}
 \end{center}
\end{figure}

The evaluation of the segment reconstruction efficiency is performed using
muon tracks reconstructed in the silicon tracker independently  of the muon chambers.
Distributions of the residuals between the reconstructed 2D $r$-$\phi$ segment
intersection with the first layer plane in MB1 and the
extrapolated tracker track position to the same plane
for the muons in four different momentum ranges (as measured by the inner tracker system) are shown in Fig.~\ref{fig:MB1extrap}.
Similar distributions are observed for chambers
MB2-MB4, with slightly increasing RMS values when going from the innermost to the outermost stations
(e.g., RMS = $8.4$~cm in MB2 and RMS = $10.7$~cm in MB4 for muons with $p_T$ in the
$[45$--$80]$~GeV/c range).
The width of the distributions is dominated by the effect of multiple scattering in the calorimeters and in
the steel return yokes of the magnet.
It decreases at larger momentum, with a behaviour well reproduced by the simulated data.
The small discrepancy at large distance values, increasing with the momentum of the muon, 
is due to fluctuations in the muon energy loss which are slightly underestimated in the simulation.
To measure the segment reconstruction efficiency, only muons with $p_T>30$~GeV/c were considered. A window of $20$~cm around the predicted position was
used to accept a segment candidate. To ensure a reliable extrapolation from the tracker tracks, when computing the efficiency for a given chamber
MBn, the extrapolation of the track to station MB(n$+1$) (exceptionally MB3 when considering the efficiency of MB4 chambers) was required to be confirmed by
a DT segment reconstructed with at least six associated hits
also in this station MB(n+1), within the same acceptance window as defined above. To avoid bias in the efficiency determination due to the trigger,
in the selection for the efficiency computation of chamber MBn it was required that the event have high-quality
local triggers delivering the same bunch crossing identification in at least two chambers in the same sector,
excluding the chamber under study. This procedure guarantees that the events were triggered independently from the trigger
response of the local trigger device of the considered chamber. The segment reconstruction efficiency as a function of the local coordinate in the chamber is shown in
 Fig.~\ref{fig:MB1eff} for the $r$-$\phi$ layers of chambers MB1-MB4 of sector 4 in YB$0$.
The observed decrease of efficiency near the chamber's edges is due to the fact that a track passing near the boundary but outside the
chamber volume can be incorrectly predicted to have its extrapolation inside the chamber.
Note that the asymmetric behaviour of the efficiency curve on the opposite sides of a chamber is 
due to the staggered geometry of the chambers in a sector (see Fig.~\ref{CMSview}, bottom part) and to the track 
selection which requires a confirmation of a good track segment, compatible with track extrapolation, 
in chamber MBn+1 when chamber MBn is under study. Due to this requirement, the chamber region near 
one edge of the chamber is not illuminated for MB1, MB2 and MB3.
 The method can be safely applied to all chambers of the three uppermost and lowermost sectors of the wheels
 YB$-1$, YB$0$ and YB$1$, where there are enough good quality tracker tracks that allow reliable extrapolation.

The DT chamber efficiency can also be evaluated making use exclusively of the information
coming from the muon spectrometer, thus extending the efficiency measurement to the chambers
of outer barrel wheels YB$\pm 2$. Muon tracks are reconstructed with the information
provided by neighbouring chambers and extrapolated to the middle of the chamber under test.
Two different approaches have been considered to reconstruct the tracks and obtain the
extrapolated position.  A simple, linear fit to the hits
recorded in the other chambers in the same sector was performed, taking into account the
uncertainty due to multiple scattering in the iron. The resulting track was extrapolated to the
chamber under test. This method is only valid for data taken with no magnetic field, as in this case muons
essentially follow a straight line trajectory.
For runs taken with the nominal magnetic field we rely on the ``Standalone Muon'' reconstruction
software~\cite{GlobalReco}.
In this case, the hits present in the chamber under test take
part in the track fitting process, thus potentially biasing the determination of the segment reconstruction
efficiency. Results obtained while applying this procedure to runs taken with zero field are,
however, compatible with those obtained from the linear tracker extrapolation.
To ensure a good accuracy in the track extrapolation and to minimize a potential bias, the following selection criteria
were applied:
\begin{itemize}
\item in both $r$-$z$ and $r$-$\phi$ planes
the number of hits associated with the muon track was required to be over 4 and 13 hits, respectively,
not counting the chamber under study;
\item  the error on the position of the extrapolation point in the chamber was required to be
smaller than 1.5~cm;
\item  the tracks must cross only a single sector and wheel;
\item  track segments in the top  (bottom) chambers of CMS are selected
if the  event was triggered on the opposite side, bottom (top) part of the detector, in order
to decouple the efficiency study from any potential trigger effects.
\end{itemize}

Most tracks with high extrapolation error have a low momentum ($p_T < 10$ GeV/c),  as they are most affected
by multiple scattering effects. Given the large amount of data recorded during CRAFT, the number of events left
after the selection is sufficient for a good efficiency measurement.
A chamber was considered to be efficient when a $r$-$\phi$ or
$r$-$z$ segment was found in that chamber within a 5~cm window around the extrapolated position (about 10 times the
RMS spread of the distribution of the spatial residuals, see next section).
The inefficiency is concentrated at chamber borders, due to
geometrical effects.  Efficiencies are higher than $99\%$ for the $r$-$\phi$ plane once
the predicted position from the extrapolation is required to be inside the chamber, at a distance
larger than 10 cm from the border (cf. Fig. \ref{fig:MB1eff}), in fair agreement with the results
obtained using tracker tracks
information for all the wheels in which the comparison between the two methods was possible.
Efficiencies in the $r$-$z$ plane are approximately $2\%$ lower,
given the smaller number of hits available for segment reconstruction in this plane.

\begin{figure}[htbp]
  \begin{center}
  \resizebox{15cm}{!}{\includegraphics{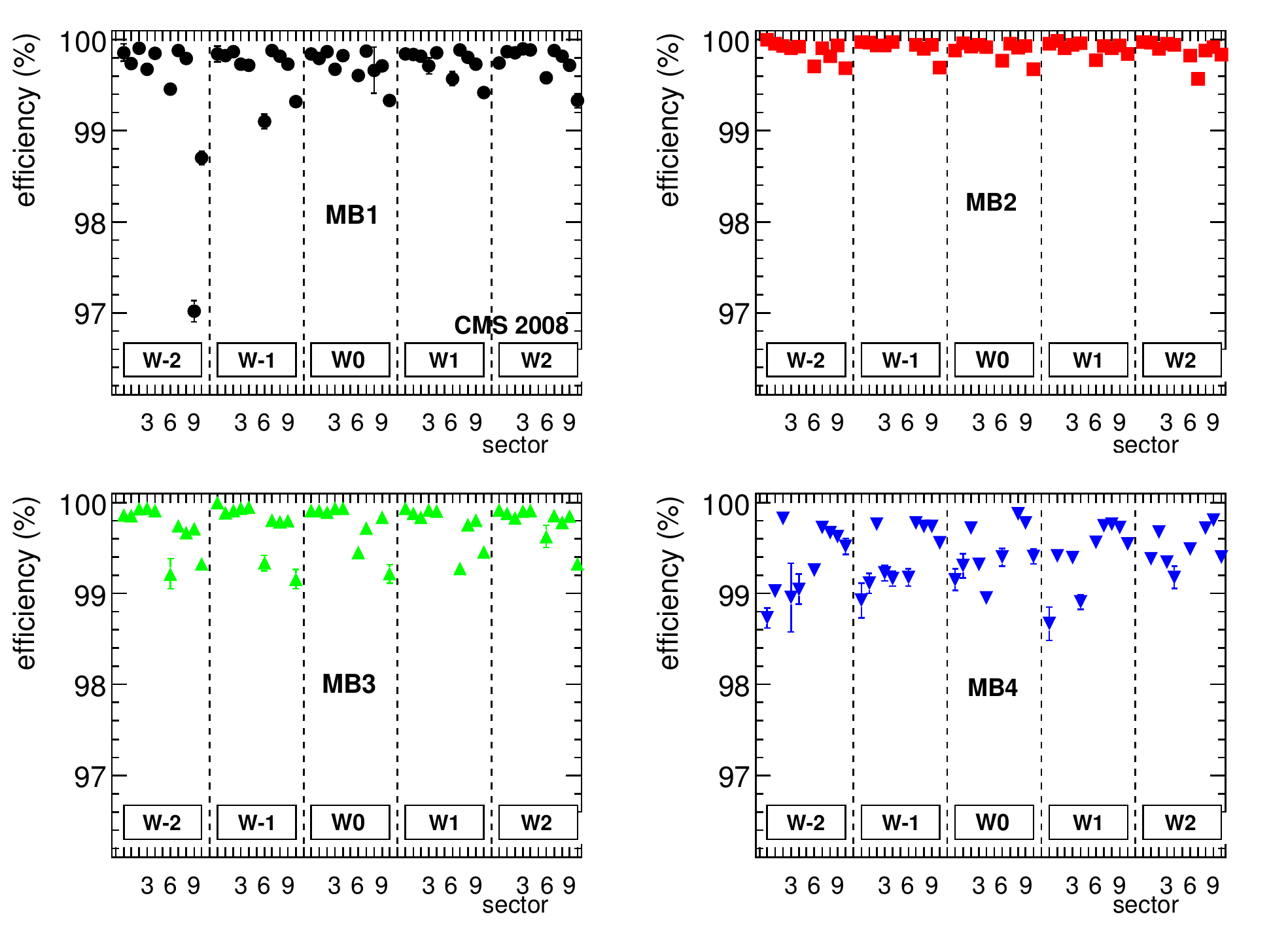}}
  \caption{ Segment reconstruction efficiency in the $r$-$\phi$ planes in the barrel muon chambers, as a function
  of the sector number for the different wheels.}
  \label{fig:effphisegments}
 \end{center}
 \end{figure}

\begin{figure}[htbp]
  \begin{center}
\resizebox{16cm}{6cm}{\includegraphics{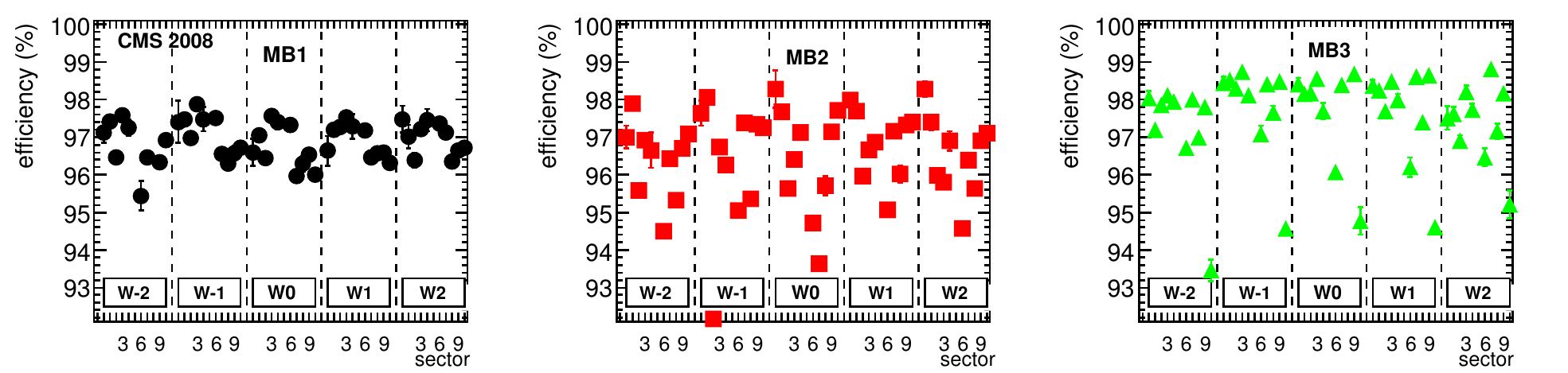}}
  \caption{  Segment reconstruction efficiency in the $r$-$z$ planes in the barrel muon chambers, as a function
  of the sector number for the different wheels.}
  \label{fig:effthetasegments}
 \end{center}
 \end{figure}

All DT sectors but the vertical ones (sectors 1 and 7) of the five wheels were
studied.
 Figure~\ref{fig:effphisegments} shows the chamber efficiency in the $r$-$\phi$ plane,
obtained by the second method described above
Every plot gives the efficiencies for a given station for each sector and wheel analyzed, marked on the horizontal axis.
Figure ~\ref{fig:effthetasegments} shows the corresponding efficiencies in the $r$-$z$ plane,
for the three chamber types (MB1/2/3) that measure the coordinate of the hit position in this plane. Results on
efficiencies are fully compatible among sectors; the drop of efficiency
observed in some of them corresponds to those sectors where the muon incident
angle is largest.
Results obtained at $B=0$~T are in agreement with those shown in Figs.~\ref{fig:effphisegments}
and ~\ref{fig:effthetasegments}.

\subsection{Track segment position and direction measurements}

\begin{figure}[htbp]
 \begin{center}
 \resizebox{7.5cm}{11cm}{\includegraphics{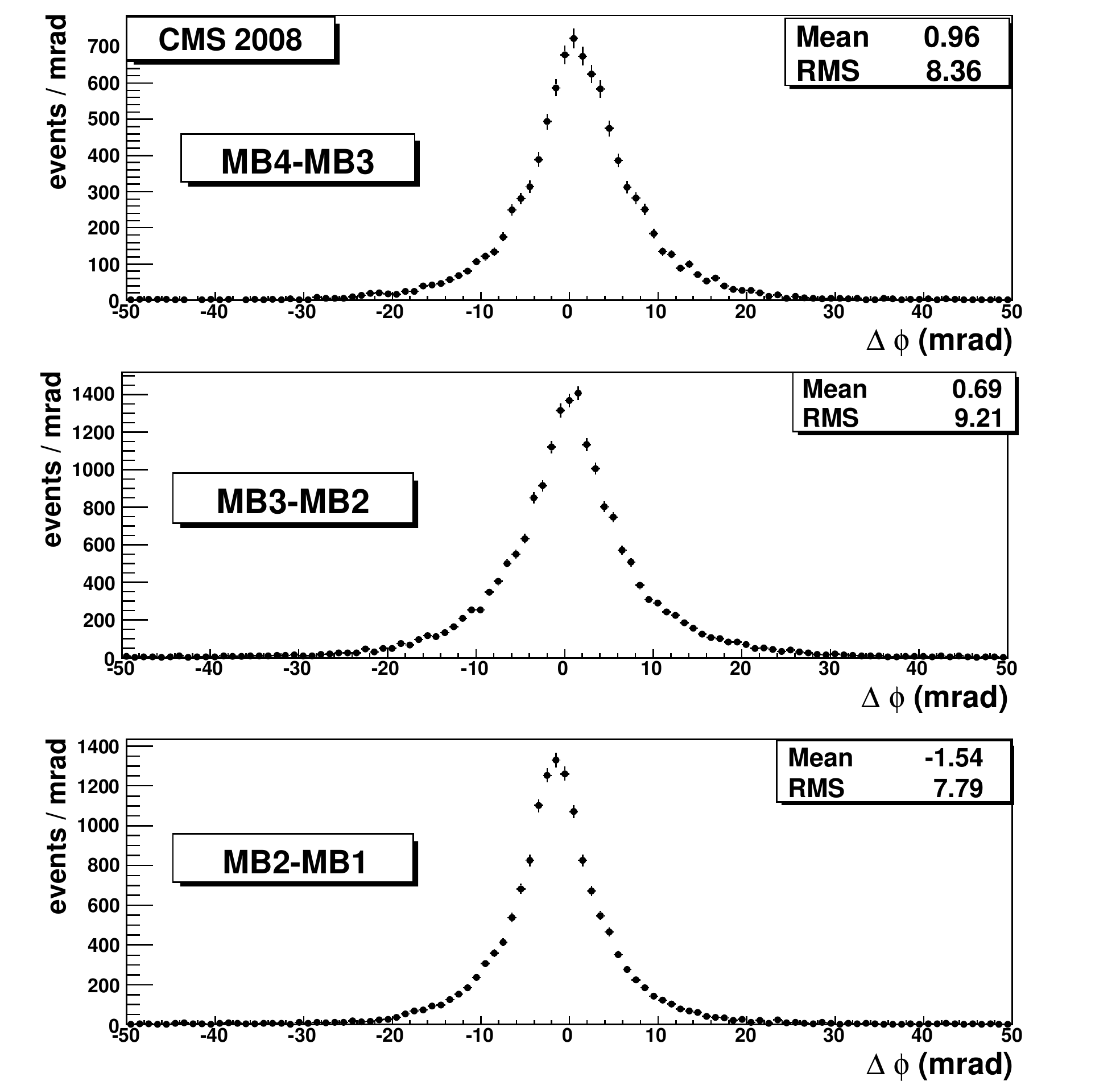}}
 \resizebox{7.5cm}{11cm}{\includegraphics{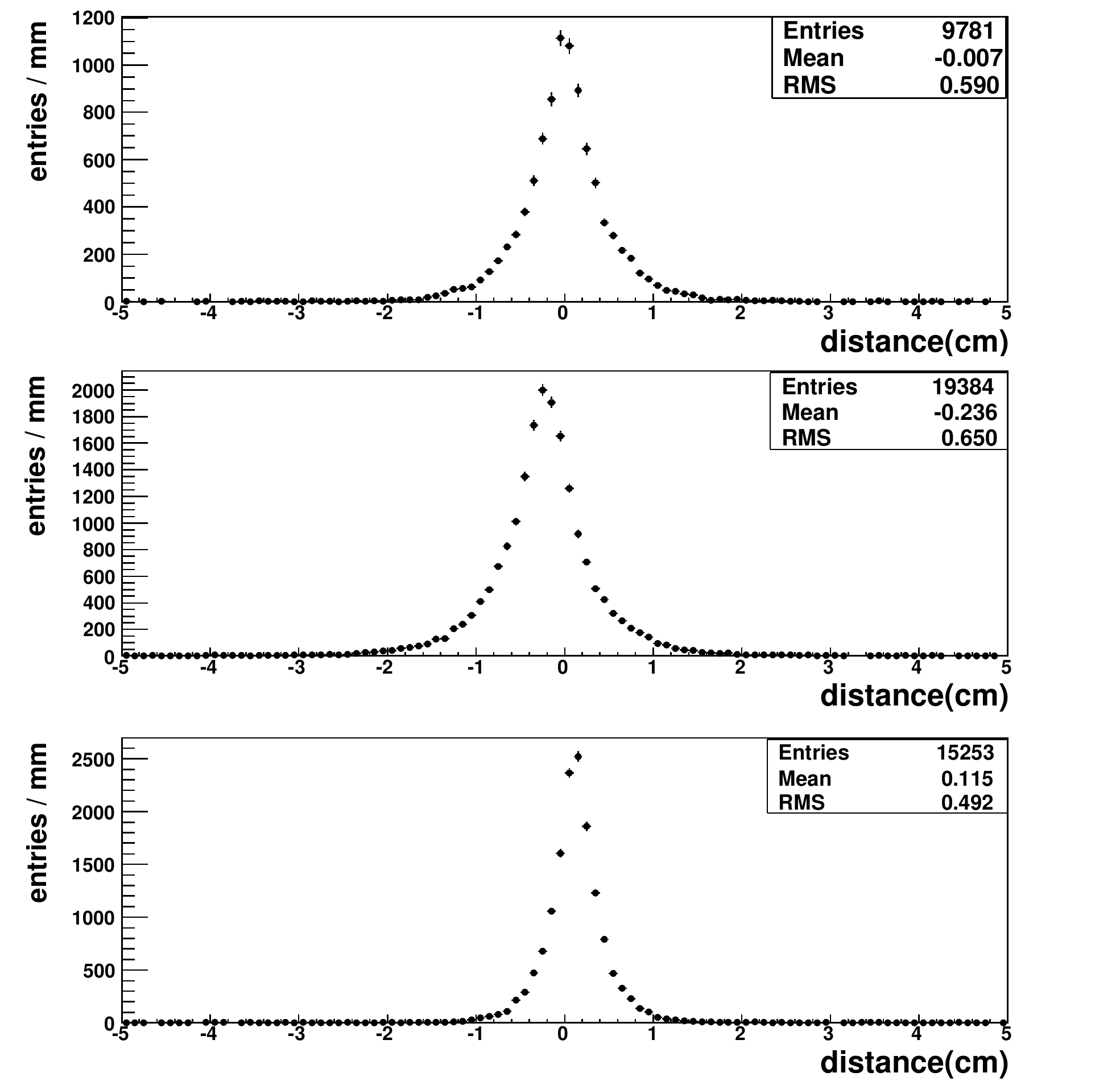}}
  \caption{ Left: distribution of the difference of the angles in the CMS bending plane of muon track
   segments reconstructed in consecutive stations in YB$-1$,
  sector 4, measured at $B=0$ T. Right: distribution of the distance between the intersection
   with the central plane of station MBn of the segment reconstructed in this station and of the
   extrapolation of the segment reconstructed in station MBn-1.}
  \label{fig:DphiB0}
 \end{center}
\end{figure}

In order to study the quality of the segment reconstruction,
the comparison of position and direction measurements
between different muon chambers for the same cosmic muon have been performed.
First, the data collected with $B=0$~T were studied. The distributions of the difference of the directions
of muon track segments reconstructed in the CMS bending plane in consecutive chambers are shown
in Fig.~\ref{fig:DphiB0} (left) for the chambers of sector 4 of YB$-1$.
The average values of the distributions are of the order of 1~mrad,
due to misalignment effects (both in the internal components of the chambers and on the relative orientation between chambers)
which are not yet completely taken into account in the reconstruction.
Figure~\ref{fig:DphiB0} (right) shows the distribution of the
distance between the intersection with the central plane of station MBn of the segment
reconstructed in this station and of the extrapolation of the segment reconstructed in MBn-1.
The average values indicate relative position misalignments between consecutive chambers of the order
of a few millimeters. It is worth noting that the smaller dispersion of the
position difference for MB1-MB2 chambers (right-bottom plot in the figure) is due to the
smaller size of the steel yoke between these chambers,
compared to the steel yokes between MB2-MB3 and MB3-MB4.

\begin{figure}[htbp]
 \begin{center}
  \resizebox{7cm}{!}{\includegraphics{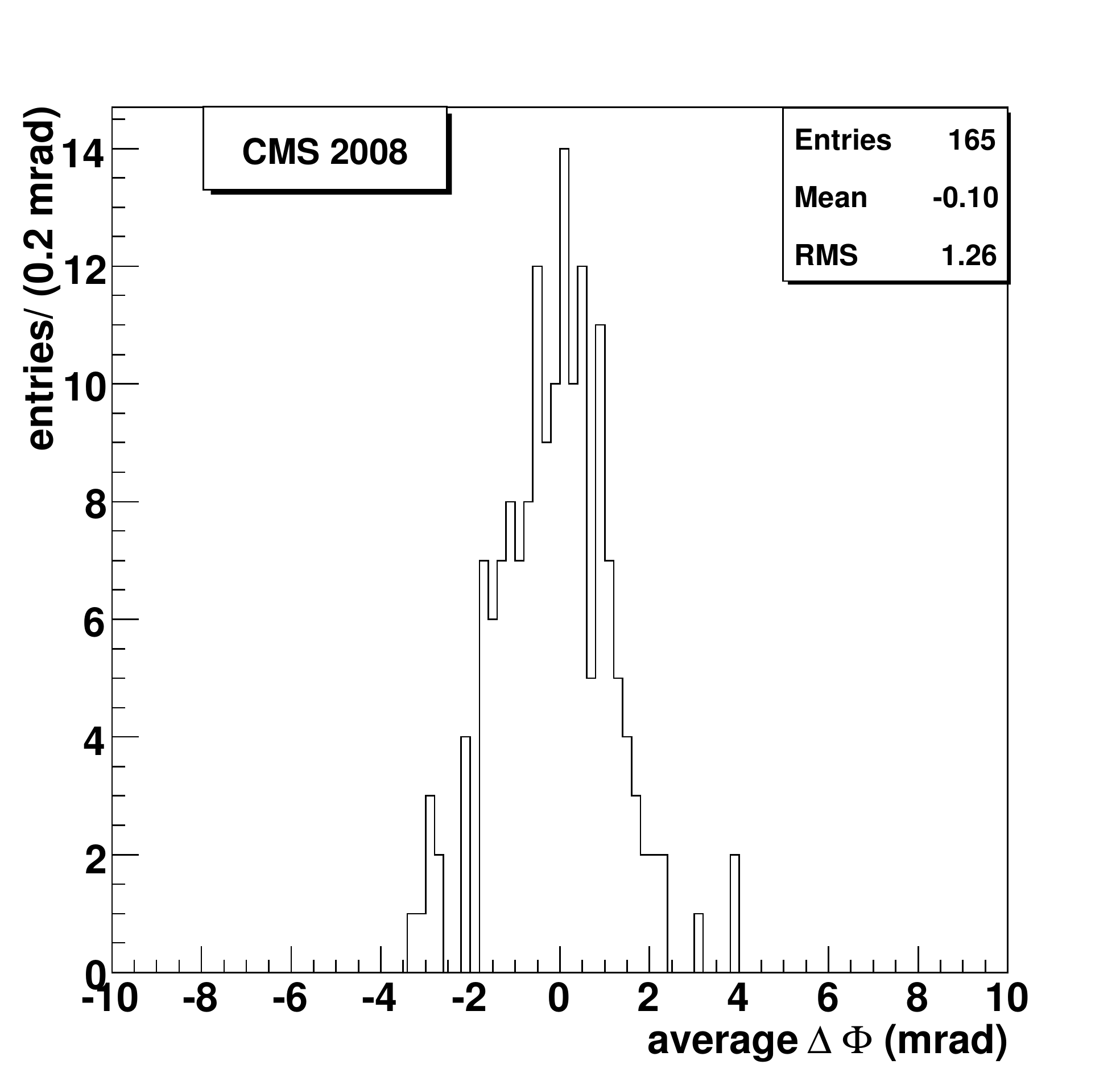}}
  \resizebox{7cm}{!}{\includegraphics{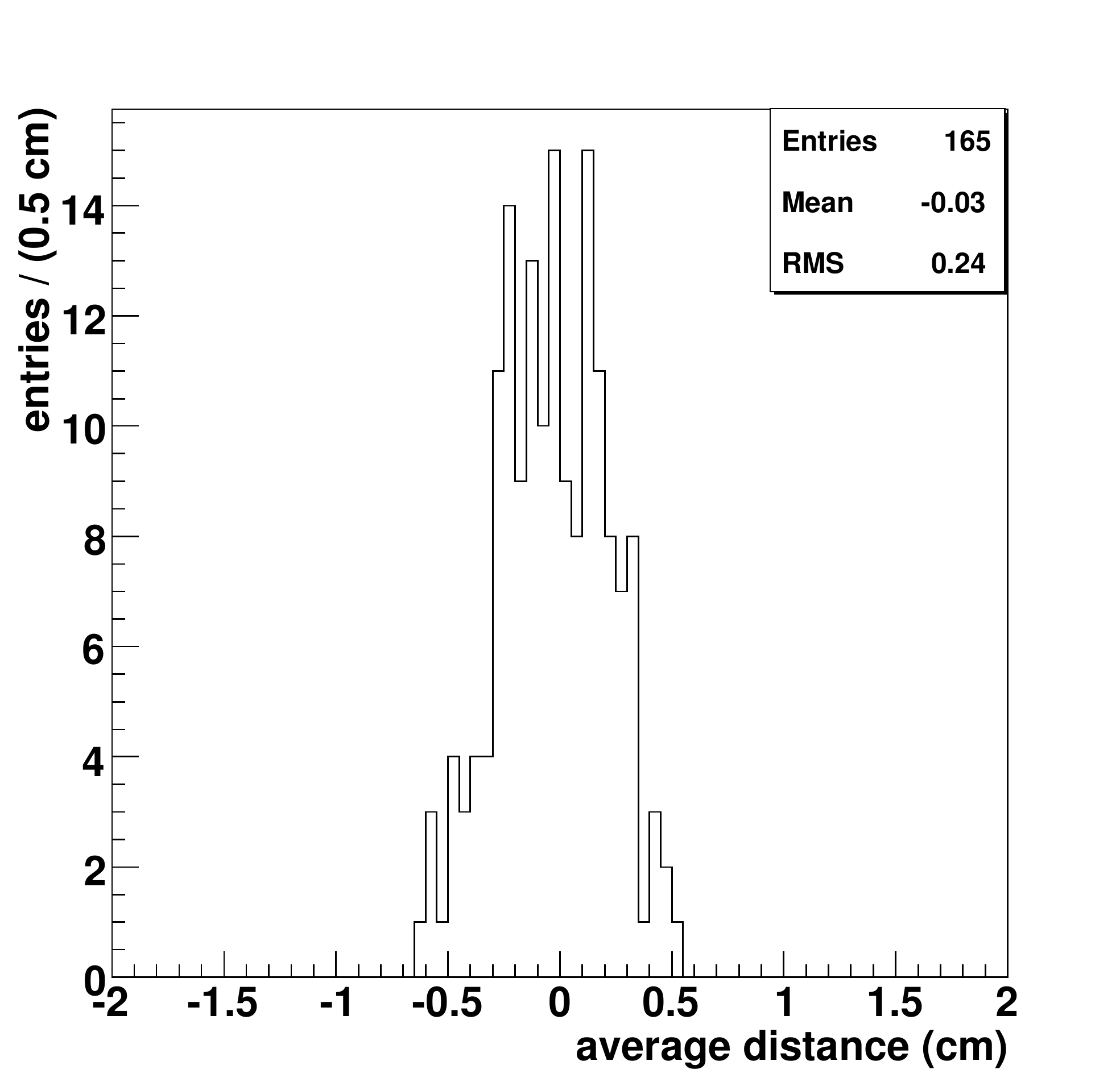}}
  \caption{ Left: distribution of the averages of the angle differences at $B=0$~T between consecutive stations in all
  wheels and sectors. Right: distribution of the averages of the distances at $B=0$~T between measured and extrapolated positions in
  consecutive chambers.}
  \label{fig:DphiB0summary}
 \end{center}
\end{figure}

The summary of the above results for all the wheels and sectors is shown in
Fig.~\ref{fig:DphiB0summary}. The distribution
of the average values of the angle differences is shown in the histogram on the left.
The RMS of the distribution is about $1$~mrad. The histogram on the right shows
a similar plot for the differences between measured and extrapolated positions.
The RMS of the distribution is about $2$~mm, showing that the relative alignment between
the chambers is compatible with the tolerance expected for the mechanical installation of the chambers in CMS.
This result guarantees that for the beginning of LHC running the muon L1 trigger processor will
operate correctly, efficiently providing muon trigger candidates with reliable estimation of their
transverse momentum. Since there is no evidence for chambers placed outside the
design mechanical tolerance, we expect that from these start-up mis-alignment conditions the use of
survey data and of the data from the laser alignment system~\cite{CRAFTalignment}
will bring the position uncertainty to the design goal of about 100~$\mu$m for
High Level Trigger and off-line reconstruction. In addition, software alignment procedures using prompt
muon tracks have been deployed, which show that it will be
possible to reach a comparable accuracy on the chambers' position after accumulating data
corresponding to a few pb$^{-1}$ of integrated luminosity~\cite{CFTalisoftw}.

\subsection{Bending power measurements}

\begin{figure}[htbp]
 \begin{center}
  \resizebox{11cm}{9cm}{\includegraphics{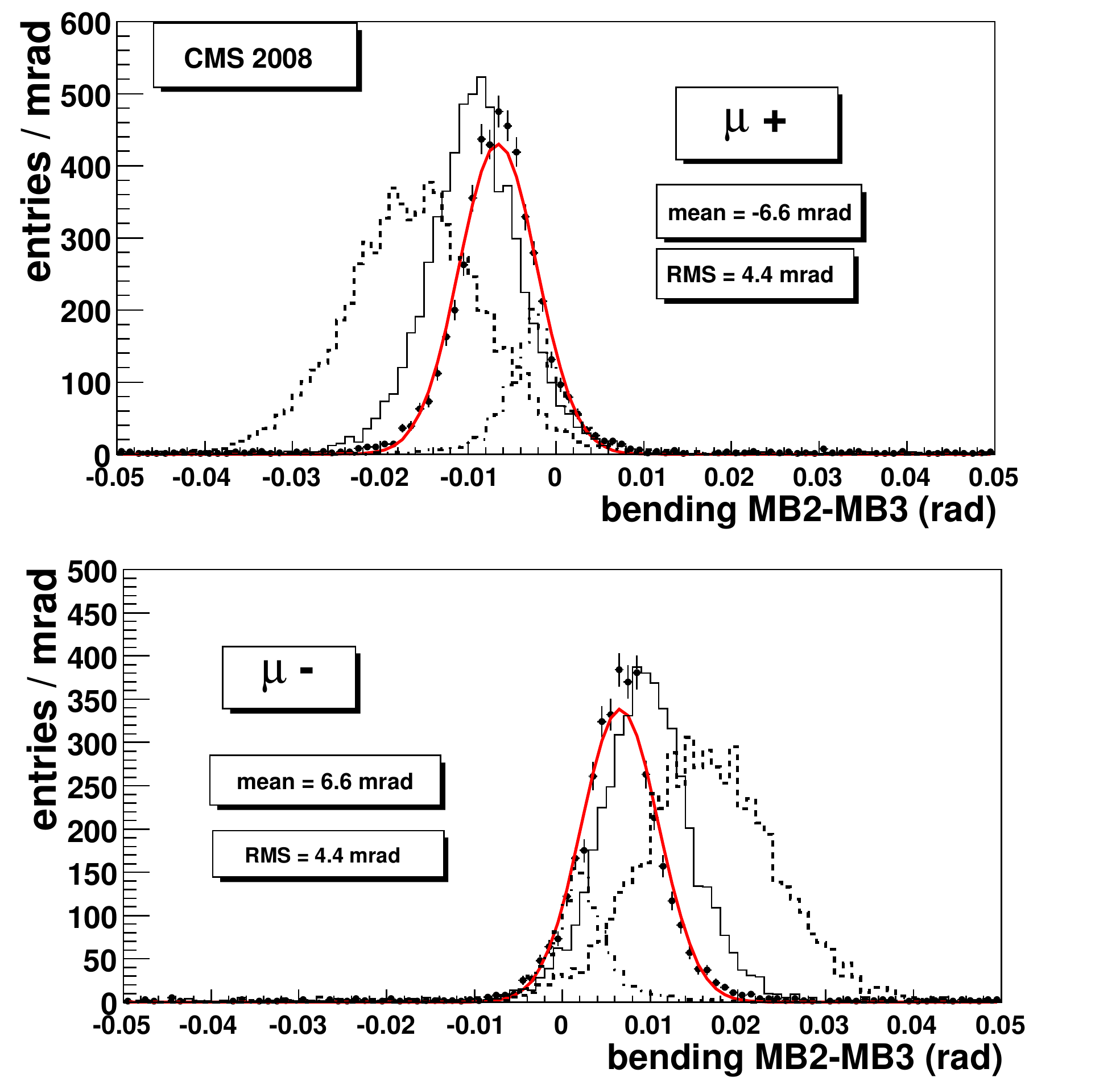}}
  \caption{ Bending angle differences between MB2 and MB3 stations. Top:  $\mu^+$; bottom:  $\mu^-$.
  Distributions for different $p_T$  intervals are shown: $[8$--$12]$ (dashed line), $[18$--$22]$ (full line),
  $[27$--$33]$ (points) and
  $[90$--$110]$ GeV/c (dashed-dotted line).
The curves show the result of a Gaussian fit to data distribution  for the $27 < p_T <33 $ GeV/c sample.}
  \label{BendingMB23}
 \end{center}
 \end{figure}

Data with the magnetic field value $B=3.8$~T in the central solenoid were used to study the bending power of the muon spectrometer.
The difference in the track angle measurements between consecutive stations were studied for different values of the transverse
momentum of the associated track, which was measured independently by the tracker. These distributions are shown
in  Fig.~\ref{BendingMB23} for MB2-MB3 pairs of stations.
As seen from the figure, the bending power for a $p_T=30$ GeV/c muon is about $6.6$~mrad. Similar distributions are observed for
MB1-MB2 and MB3-MB4 pairs of stations, with bending power equal to $4.0$~mrad  and  $6.0$~mrad, respectively.
Note that the width of the magnetized steel between the chambers is about $30$~cm
between MB1 and MB2 and $62$~cm between MB2-MB3 and MB3-MB4~\cite{MagnetTDR}. The magnetic flux density in the
steel yokes decreases slightly with the radial position.

\begin{figure}[htbp]
 \begin{center}
\resizebox{11cm}{9cm}{\includegraphics{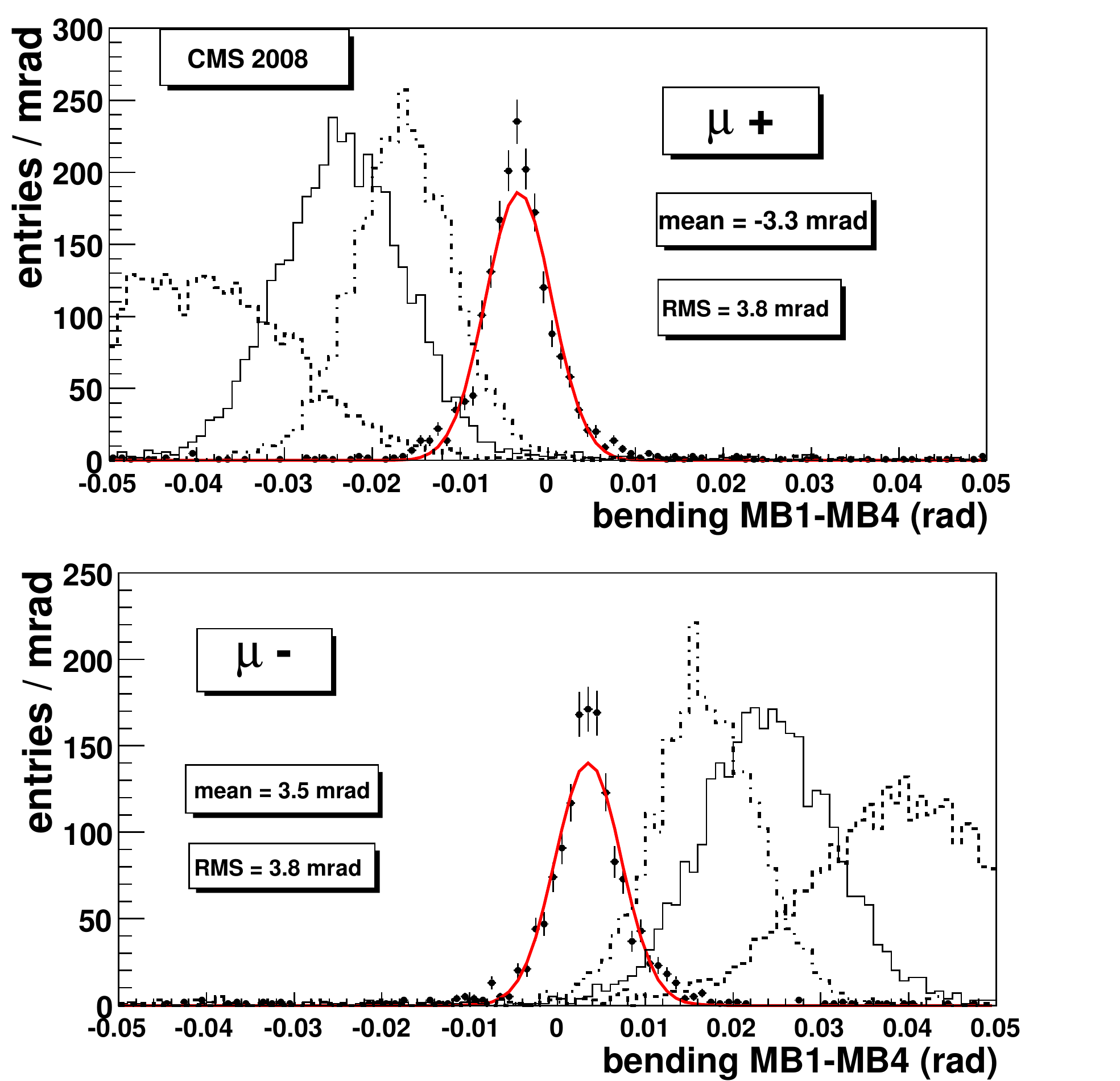}}
  \caption{ Bending angle differences between MB1 and MB4 stations. Top:  $\mu^+$; bottom:  $\mu^-$.
  Distributions for different $p_T$  intervals are shown: $[8$--$12]$ (dashed line),
  $[18$--$22]$ (full line), $[27$--$33]$ (dashed-dotted line) and
  $[150$--$250]$ GeV/c (points).
The curves show the result of a Gaussian fit to data distribution  for the $150 < p_T <250 $ GeV/c sample.}
  \label{BendingMB14}
 \end{center}
 \end{figure}

  Figure~\ref{BendingMB14} shows the distributions of the angle difference between MB1 and MB4 stations, displaying the bending power of the full
  lever arm in a barrel sector. For muons selected in the $p_T$ range $[150,250]$ GeV/c, the average deflection
  by the magnetic field in the steel return yokes of the magnet is about $3.4$~mrad.


\begin{figure}[htbp]
 \begin{center}
  \resizebox{7.5cm}{!}{\includegraphics{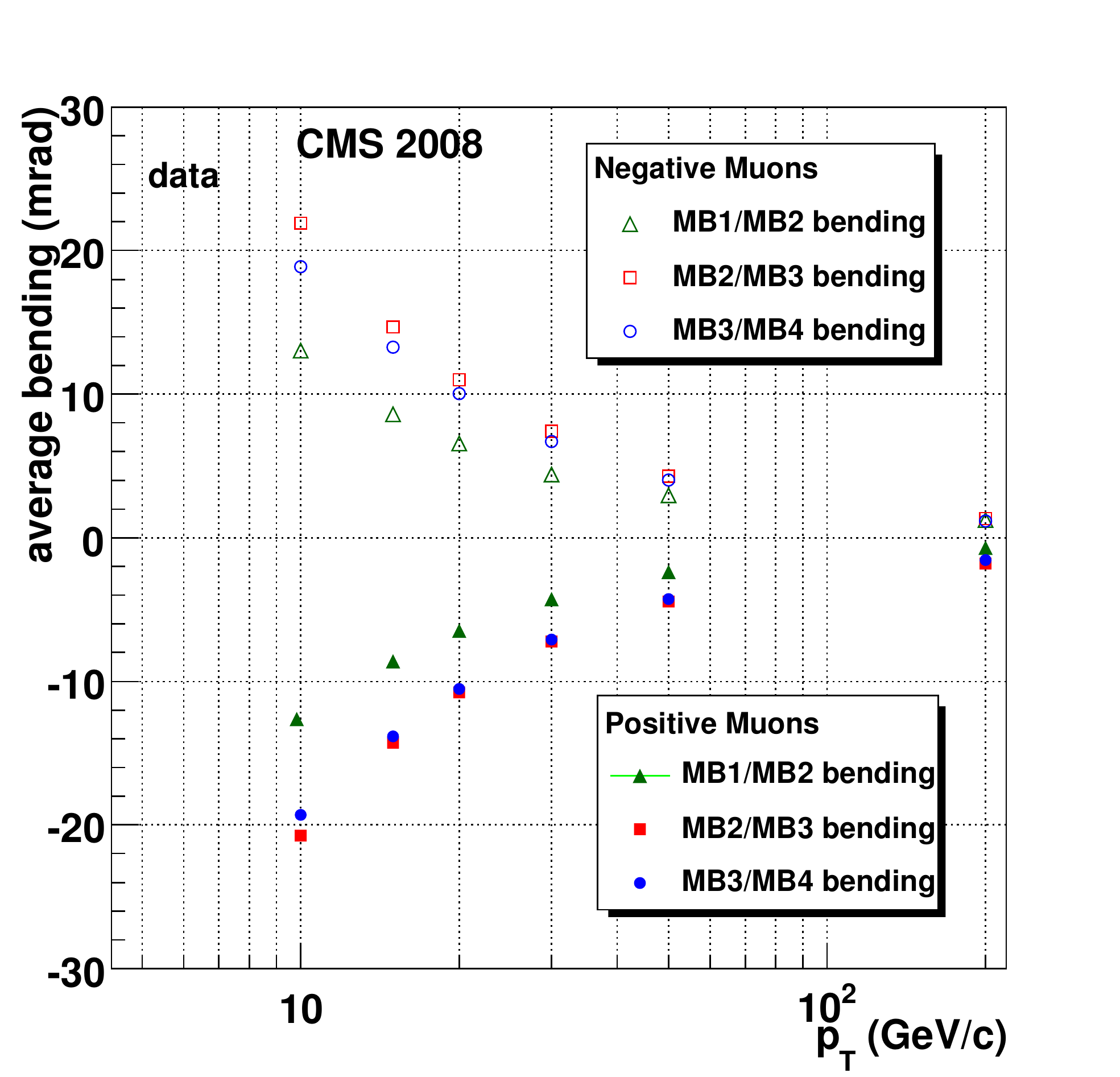}}
  \resizebox{7.5cm}{!}{\includegraphics{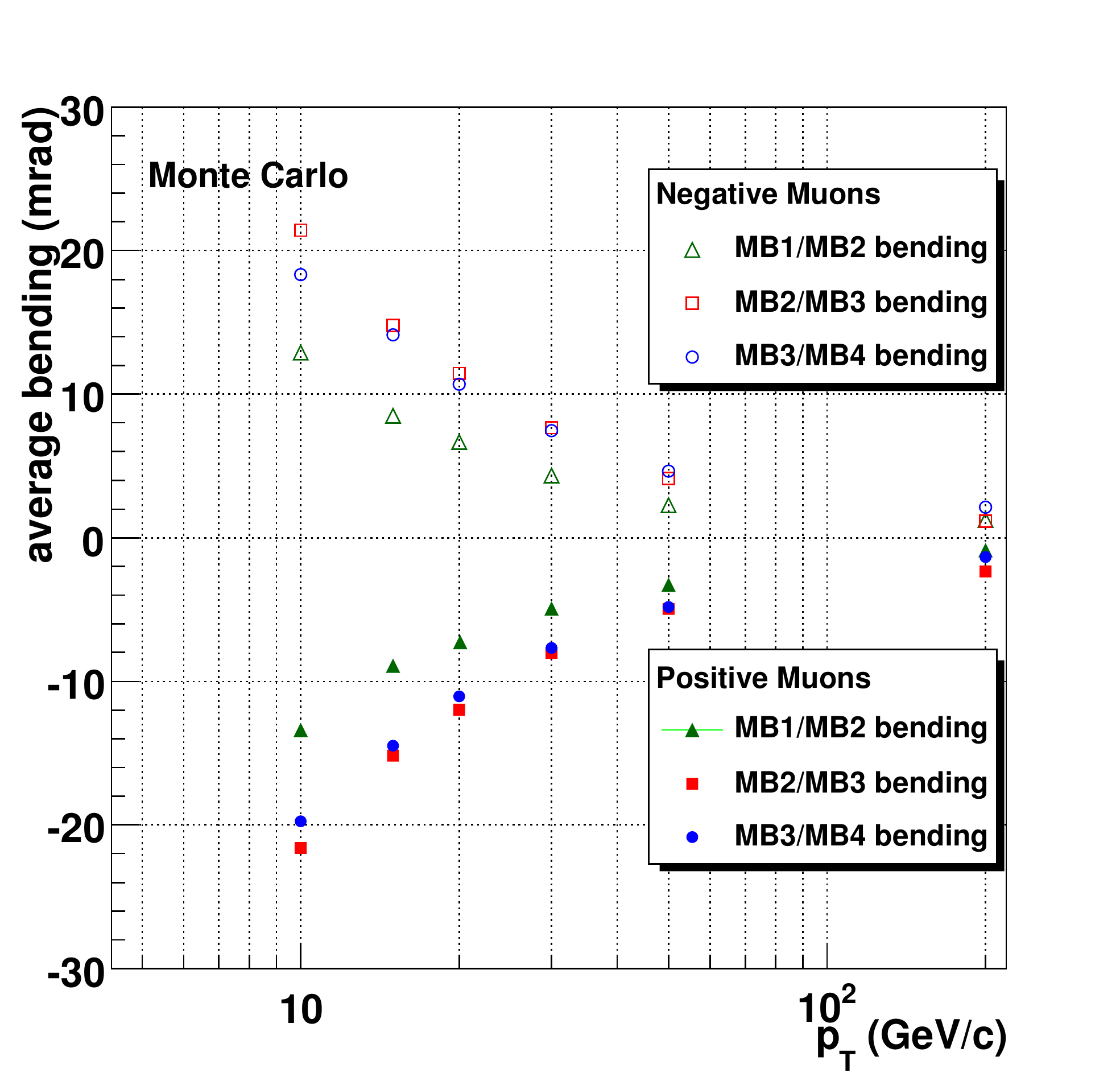}}
  \caption{ Mean values of Gaussian fits to the distributions of the bending angle differences between consecutive stations as a function of the muon $p_T$.
  Data are for the magnetic field value $B=3.8$ T in the central solenoid. Left: real data; right: simulated data.}
  \label{BendingvsPt}
 \end{center}
 \end{figure}

\begin{figure}[htbp]
 \begin{center}
\resizebox{9cm}{!}{\includegraphics{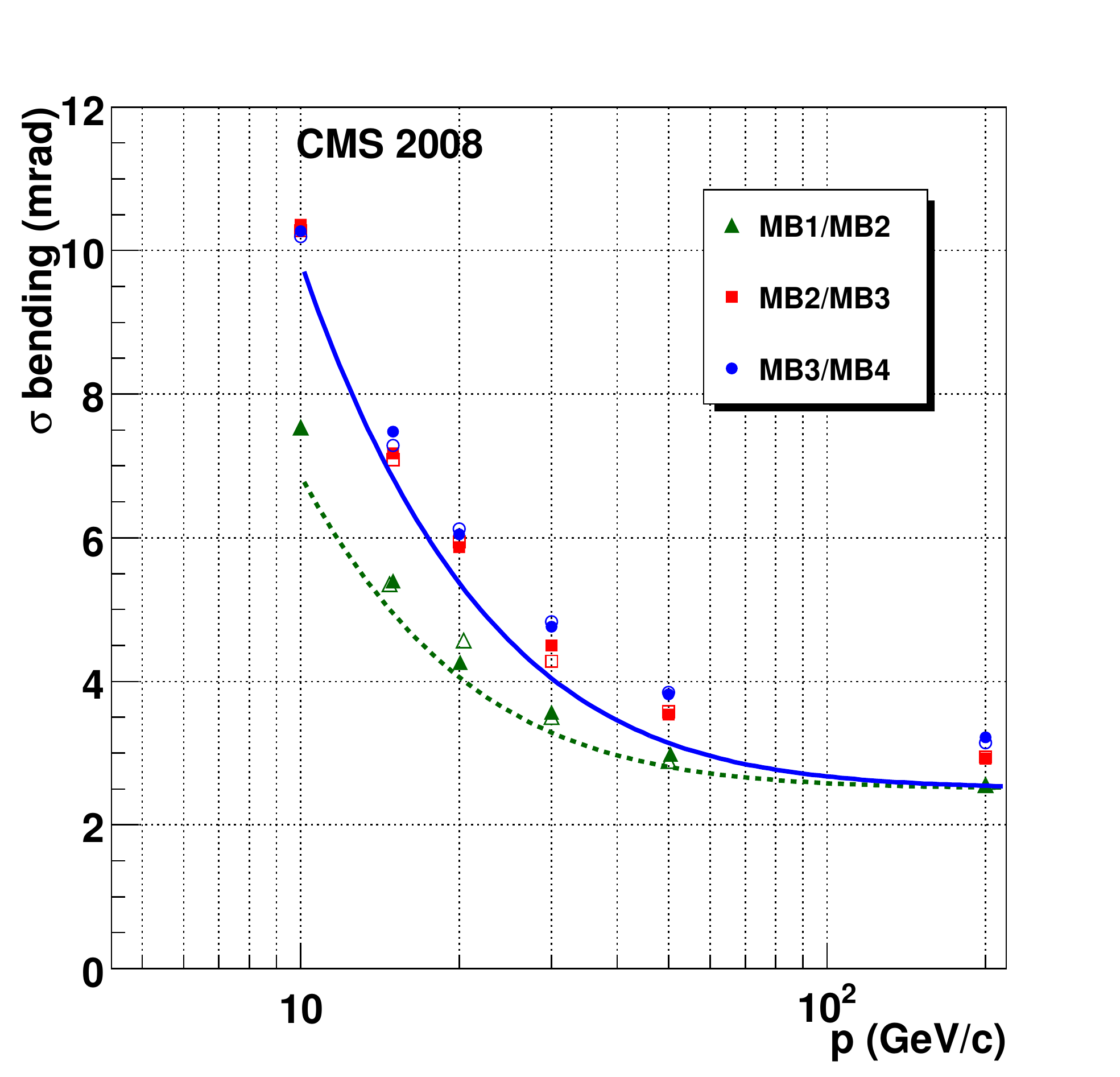}}
  \caption{Widths of Gaussian fits to the distributions of the bending angle differences between consecutive
  stations as a function of the muon momentum.
  Full points: $\mu^+$; open points: $\mu^-$.
  The dashed (full line) curves show the expectation from multiple scattering between MB1-MB2(MB2-MB3 and MB3-MB4) stations,
  with $2.5$~mrad added in quadrature, corresponding to the asymptotic value extracted at high momenta.}
  \label{SigmaBendingvsP}
 \end{center}
 \end{figure}

The above results and a comparison with the simulation are summarized in Figs.~\ref{BendingvsPt}
and ~\ref{SigmaBendingvsP}, where the average and the width
of the Gaussian fits to the distributions of the angle difference are plotted versus the transverse momentum
of the track. The results are shown both for positive (full points) and negative muons (open points).
The behaviour shown in Fig.~\ref{SigmaBendingvsP} is consistent with the expectations
from the multiple scattering in the iron:
the dashed and full lines show respectively the computation for an average material crossed
by the muon between
the two innermost stations of MB1 and MB2 (18 radiation lengths) and of MB2 and MB3 or MB3 and MB4
(37 radiation lengths)~\cite{MUtdr}, summed in quadrature with a constant term 
$\sigma_\propto = 2.5$~mrad. This asymptotical value reached at high momenta is compatible with expectations.
In fact, the intrinsic angular resolution of each chamber expected from the observed single hit 
resolution (on average about 290 $\mu$m, see Fig.~\ref{AllFitRes}, for the core of the hit residual distribution) 
is $\sigma_{intrins.} =1.5$~mrad, taking into account the hit multiplicity distributions of the 
segments and the presence of tails in the hit residual distributions shown in Fig.~\ref{HitRes}. 
The contribution to the bending measurement error from the incomplete knowledge of chambers' alignment,
as extracted from $B=0$ T data (cf. the distribution of the $\Delta \phi$  averages shown 
in Fig.~\ref{fig:DphiB0summary}) 
is $\sigma_{mis-align} = 1.3$ mrad$/ \sqrt{2} = 0.9$ mrad. The expected value for $\sigma_\propto$ is thus 
 $\sigma_\propto  = [ \sigma_{intrins.}^2 + \sigma_{mis-align}^2]^{1/2} \cdot \sqrt{2} = 2.5$~mrad.  
It must be noted that the asymptotic  behaviour for MB2-MB3 and MB3-MB4 is slightly worse, mainly due to the 
fact that the modelling  of the multiple scattering effects with a Gaussian curve tends to be 
inadequate when increasing the amount of material crossed by the muon. 
The intrinsic angular resolution measured in dedicated bunched test beams~\cite{tbeam2004} was about $1$~mrad
for muon tracks normal to the chamber plane.
It is worth stressing here that the present result is obtained using segment tracks with a very
large angular spread with respect to the direction normal to the chambers' plane,
for which the chamber behaviour is optimal.
The angular spread of segments originating from prompt muons produced in pp collisions is considerably smaller.

\section{Conclusions}

The performance of the DT barrel muon detector of CMS was studied in detail using cosmic muon data
collected in autumn 2008, both with zero magnetic  field  and with the magnet solenoid operating at $B=3.8$~T. The data analysis performed
on 246 out of the total of 250 DT chambers shows a very good muon reconstruction capability, with a resolution of single reconstructed hits
on the order of $260$~$\mu$m in all chambers except the vertical ones, which
could not be studied well with cosmic rays.
The reconstruction efficiency of  high-quality
local track segments in each station has been measured to be about $99\%$ in all chambers.
The comparison between measurements of the track segment positions and directions in the different
chambers shows a behaviour compatible
with the expectations from the multiple scattering of the muons in the steel yoke.
The spread in the measurement of the track direction in the bending plane of CMS
was about $6$~mrad, averaged over the whole momentum spectrum of cosmic muons
with $p_T > 10$ GeV/c. The bending power in the steel return yoke between
the innermost and outermost station has been measured to be about $3$~mrad for $p_T= 200$~GeV/c muons.
The relative misalignments of the chambers, as measured by the data collected at $B=0$~T,
are well within the mechanical tolerances (a few mm) for the insertion
of the chambers into their cradles inside the magnet yoke structure.

The chamber performance is in good agreement with the simulation; it provides a good starting point
that assures fully efficient operation of the muon DT trigger and eventual achievement of
the original design criteria of the DT system. The criteria specify robust and efficient muon identification, and the capability
of measuring the muon position
in each station with a precision of about $100$~$\mu$m, in order to provide good momentum resolution for
highly energetic muons.
The above results are very encouraging and allow the anticipation of a good performance of the DT
barrel muon detector during early phases of
LHC operation and data taking, which would provide efficient identification and reconstruction of muons.

\section*{Acknowledgements}


We thank the technical and administrative staff at CERN and other CMS Institutes,
and acknowledge support from: FMSR (Austria); FNRS and FWO (Belgium);
CNPq, CAPES, FAPERJ, and FAPESP (Brazil); MES (Bulgaria); CERN; CAS, MoST,
and NSFC (China); COLCIENCIAS (Colombia); MSES (Croatia); RPF (Cyprus);
Academy of Sciences and NICPB (Estonia); Academy of Finland, ME, and HIP (Finland);
CEA and CNRS/IN2P3 (France); BMBF, DFG, and HGF (Germany); GSRT (Greece);
OTKA and NKTH (Hungary); DAE and DST (India); IPM (Iran); SFI (Ireland);
INFN (Italy); NRF (Korea); LAS (Lithuania); CINVESTAV, CONACYT, SEP, and
UASLP-FAI (Mexico); PAEC (Pakistan); SCSR (Poland); FCT (Portugal); JINR (Armenia,
Belarus, Georgia, Ukraine, Uzbekistan); MST and MAE (Russia); MSTDS (Serbia);
MICINN and CPAN (Spain); Swiss Funding Agencies (Switzerland); NSC (Taipei);
TUBITAK and TAEK (Turkey); STFC (United Kingdom); DOE and NSF (USA).
Individuals have received support from the Marie-Curie IEF program (European Union);
the Leventis Foundation; the A. P. Sloan Foundation;
and the Alexander von Humboldt Foundation.

\bibliography{auto_generated}   

\cleardoublepage\appendix\section{The CMS Collaboration \label{app:collab}}\begin{sloppypar}\hyphenpenalty=500\input{CFT-09-012-authorlist.tex}\end{sloppypar}
\end{document}

%% file: ptdr-definitions.tex
%
%
%

\newcommand {\etal}{\mbox{et al.}\xspace} 
\newcommand {\ie}{\mbox{i.e.}\xspace}     
\newcommand {\eg}{\mbox{e.g.}\xspace}     
\newcommand {\etc}{\mbox{etc.}\xspace}     
\newcommand {\vs}{\mbox{\sl vs.}\xspace}      
\newcommand {\mdash}{\ensuremath{\mathrm{-}}} 

\newcommand {\Lone}{Level-1\xspace} 
\newcommand {\Ltwo}{Level-2\xspace}
\newcommand {\Lthree}{Level-3\xspace}

\providecommand{\ACERMC} {\textsc{AcerMC}\xspace}
\providecommand{\ALPGEN} {{\textsc{alpgen}}\xspace}
\providecommand{\CHARYBDIS} {{\textsc{charybdis}}\xspace}
\providecommand{\CMKIN} {\textsc{cmkin}\xspace}
\providecommand{\CMSIM} {{\textsc{cmsim}}\xspace}
\providecommand{\CMSSW} {{\textsc{cmssw}}\xspace}
\providecommand{\COBRA} {{\textsc{cobra}}\xspace}
\providecommand{\COCOA} {{\textsc{cocoa}}\xspace}
\providecommand{\COMPHEP} {\textsc{CompHEP}\xspace}
\providecommand{\EVTGEN} {{\textsc{evtgen}}\xspace}
\providecommand{\FAMOS} {{\textsc{famos}}\xspace}
\providecommand{\GARCON} {\textsc{garcon}\xspace}
\providecommand{\GARFIELD} {{\textsc{garfield}}\xspace}
\providecommand{\GEANE} {{\textsc{geane}}\xspace}
\providecommand{\GEANTfour} {{\textsc{geant4}}\xspace}
\providecommand{\GEANTthree} {{\textsc{geant3}}\xspace}
\providecommand{\GEANT} {{\textsc{geant}}\xspace}
\providecommand{\HDECAY} {\textsc{hdecay}\xspace}
\providecommand{\HERWIG} {{\textsc{herwig}}\xspace}
\providecommand{\HIGLU} {{\textsc{higlu}}\xspace}
\providecommand{\HIJING} {{\textsc{hijing}}\xspace}
\providecommand{\IGUANA} {\textsc{iguana}\xspace}
\providecommand{\ISAJET} {{\textsc{isajet}}\xspace}
\providecommand{\ISAPYTHIA} {{\textsc{isapythia}}\xspace}
\providecommand{\ISASUGRA} {{\textsc{isasugra}}\xspace}
\providecommand{\ISASUSY} {{\textsc{isasusy}}\xspace}
\providecommand{\ISAWIG} {{\textsc{isawig}}\xspace}
\providecommand{\MADGRAPH} {\textsc{MadGraph}\xspace}
\providecommand{\MCATNLO} {\textsc{mc@nlo}\xspace}
\providecommand{\MCFM} {\textsc{mcfm}\xspace}
\providecommand{\MILLEPEDE} {{\textsc{millepede}}\xspace}
\providecommand{\ORCA} {{\textsc{orca}}\xspace}
\providecommand{\OSCAR} {{\textsc{oscar}}\xspace}
\providecommand{\PHOTOS} {\textsc{photos}\xspace}
\providecommand{\PROSPINO} {\textsc{prospino}\xspace}
\providecommand{\PYTHIA} {{\textsc{pythia}}\xspace}
\providecommand{\SHERPA} {{\textsc{sherpa}}\xspace}
\providecommand{\TAUOLA} {\textsc{tauola}\xspace}
\providecommand{\TOPREX} {\textsc{TopReX}\xspace}
\providecommand{\XDAQ} {{\textsc{xdaq}}\xspace}

\newcommand {\DZERO}{D\O\xspace}     


\newcommand{\de}{\ensuremath{^\circ}}
\newcommand{\ten}[1]{\ensuremath{\times \text{10}^\text{#1}}}
\newcommand{\unit}[1]{\ensuremath{\text{\,#1}}\xspace}
\newcommand{\mum}{\ensuremath{\,\mu\text{m}}\xspace}
\newcommand{\micron}{\ensuremath{\,\mu\text{m}}\xspace}
\newcommand{\cm}{\ensuremath{\,\text{cm}}\xspace}
\newcommand{\mm}{\ensuremath{\,\text{mm}}\xspace}
\newcommand{\mus}{\ensuremath{\,\mu\text{s}}\xspace}
\newcommand{\keV}{\ensuremath{\,\text{ke\hspace{-.08em}V}}\xspace}
\newcommand{\MeV}{\ensuremath{\,\text{Me\hspace{-.08em}V}}\xspace}
\newcommand{\GeV}{\ensuremath{\,\text{Ge\hspace{-.08em}V}}\xspace}
\newcommand{\TeV}{\ensuremath{\,\text{Te\hspace{-.08em}V}}\xspace}
\newcommand{\PeV}{\ensuremath{\,\text{Pe\hspace{-.08em}V}}\xspace}
\newcommand{\keVc}{\ensuremath{{\,\text{ke\hspace{-.08em}V\hspace{-0.16em}/\hspace{-0.08em}c}}}\xspace}
\newcommand{\MeVc}{\ensuremath{{\,\text{Me\hspace{-.08em}V\hspace{-0.16em}/\hspace{-0.08em}c}}}\xspace}
\newcommand{\GeVc}{\ensuremath{{\,\text{Ge\hspace{-.08em}V\hspace{-0.16em}/\hspace{-0.08em}c}}}\xspace}
\newcommand{\TeVc}{\ensuremath{{\,\text{Te\hspace{-.08em}V\hspace{-0.16em}/\hspace{-0.08em}c}}}\xspace}
\newcommand{\keVcc}{\ensuremath{{\,\text{ke\hspace{-.08em}V\hspace{-0.16em}/\hspace{-0.08em}c}^\text{2}}}\xspace}
\newcommand{\MeVcc}{\ensuremath{{\,\text{Me\hspace{-.08em}V\hspace{-0.16em}/\hspace{-0.08em}c}^\text{2}}}\xspace}
\newcommand{\GeVcc}{\ensuremath{{\,\text{Ge\hspace{-.08em}V\hspace{-0.16em}/\hspace{-0.08em}c}^\text{2}}}\xspace}
\newcommand{\TeVcc}{\ensuremath{{\,\text{Te\hspace{-.08em}V\hspace{-0.16em}/\hspace{-0.08em}c}^\text{2}}}\xspace}

\newcommand{\pbinv} {\mbox{\ensuremath{\,\text{pb}^\text{$-$1}}}\xspace}
\newcommand{\fbinv} {\mbox{\ensuremath{\,\text{fb}^\text{$-$1}}}\xspace}
\newcommand{\nbinv} {\mbox{\ensuremath{\,\text{nb}^\text{$-$1}}}\xspace}
\newcommand{\percms}{\ensuremath{\,\text{cm}^\text{$-$2}\,\text{s}^\text{$-$1}}\xspace}
\newcommand{\lumi}{\ensuremath{\mathcal{L}}\xspace}
\newcommand{\Lumi}{\ensuremath{\mathcal{L}}\xspace}
%
\newcommand{\LvLow}  {\ensuremath{\mathcal{L}=\text{10}^\text{32}\,\text{cm}^\text{$-$2}\,\text{s}^\text{$-$1}}\xspace}
\newcommand{\LLow}   {\ensuremath{\mathcal{L}=\text{10}^\text{33}\,\text{cm}^\text{$-$2}\,\text{s}^\text{$-$1}}\xspace}
\newcommand{\lowlumi}{\ensuremath{\mathcal{L}=\text{2}\times \text{10}^\text{33}\,\text{cm}^\text{$-$2}\,\text{s}^\text{$-$1}}\xspace}
\newcommand{\LMed}   {\ensuremath{\mathcal{L}=\text{2}\times \text{10}^\text{33}\,\text{cm}^\text{$-$2}\,\text{s}^\text{$-$1}}\xspace}
\newcommand{\LHigh}  {\ensuremath{\mathcal{L}=\text{10}^\text{34}\,\text{cm}^\text{$-$2}\,\text{s}^\text{$-$1}}\xspace}
\newcommand{\hilumi} {\ensuremath{\mathcal{L}=\text{10}^\text{34}\,\text{cm}^\text{$-$2}\,\text{s}^\text{$-$1}}\xspace}


\newcommand{\zp}{\ensuremath{\mathrm{Z}^\prime}\xspace}


\newcommand{\kt}{\ensuremath{k_{\mathrm{T}}}\xspace}
\newcommand{\BC}{\ensuremath{{B_{\mathrm{c}}}}\xspace}
\newcommand{\bbarc}{\ensuremath{{\overline{b}c}}\xspace}
\newcommand{\bbbar}{\ensuremath{{b\overline{b}}}\xspace}
\newcommand{\ccbar}{\ensuremath{{c\overline{c}}}\xspace}
\newcommand{\JPsi}{\ensuremath{{J}/\psi}\xspace}
\newcommand{\bspsiphi}{\ensuremath{B_s \to \JPsi\, \phi}\xspace}
\newcommand{\AFB}{\ensuremath{A_\mathrm{FB}}\xspace}
\newcommand{\EE}{\ensuremath{e^+e^-}\xspace}
\newcommand{\MM}{\ensuremath{\mu^+\mu^-}\xspace}
\newcommand{\TT}{\ensuremath{\tau^+\tau^-}\xspace}
\newcommand{\wangle}{\ensuremath{\sin^{2}\theta_{\mathrm{eff}}^\mathrm{lept}(M^2_\mathrm{Z})}\xspace}
\newcommand{\ttbar}{\ensuremath{{t\overline{t}}}\xspace}
\newcommand{\stat}{\ensuremath{\,\text{(stat.)}}\xspace}
\newcommand{\syst}{\ensuremath{\,\text{(syst.)}}\xspace}

\newcommand{\HGG}{\ensuremath{\mathrm{H}\to\gamma\gamma}}
\newcommand{\gev}{\GeV}
\newcommand{\GAMJET}{\ensuremath{\gamma + \mathrm{jet}}}
\newcommand{\PPTOJETS}{\ensuremath{\mathrm{pp}\to\mathrm{jets}}}
\newcommand{\PPTOGG}{\ensuremath{\mathrm{pp}\to\gamma\gamma}}
\newcommand{\PPTOGAMJET}{\ensuremath{\mathrm{pp}\to\gamma +
\mathrm{jet}
}}
\newcommand{\MH}{\ensuremath{\mathrm{M_{\mathrm{H}}}}}
\newcommand{\RNINE}{\ensuremath{\mathrm{R}_\mathrm{9}}}
\newcommand{\DR}{\ensuremath{\Delta\mathrm{R}}}


\newcommand{\PT}{\ensuremath{p_{\mathrm{T}}}\xspace}
\newcommand{\pt}{\ensuremath{p_{\mathrm{T}}}\xspace}
\newcommand{\ET}{\ensuremath{E_{\mathrm{T}}}\xspace}
\newcommand{\HT}{\ensuremath{H_{\mathrm{T}}}\xspace}
\newcommand{\et}{\ensuremath{E_{\mathrm{T}}}\xspace}
\newcommand{\Em}{\ensuremath{E\!\!\!/}\xspace}
\newcommand{\Pm}{\ensuremath{p\!\!\!/}\xspace}
\newcommand{\PTm}{\ensuremath{{p\!\!\!/}_{\mathrm{T}}}\xspace}
\newcommand{\ETm}{\ensuremath{E_{\mathrm{T}}^{\mathrm{miss}}}\xspace}
\newcommand{\MET}{\ensuremath{E_{\mathrm{T}}^{\mathrm{miss}}}\xspace}
\newcommand{\ETmiss}{\ensuremath{E_{\mathrm{T}}^{\mathrm{miss}}}\xspace}
\newcommand{\VEtmiss}{\ensuremath{{\vec E}_{\mathrm{T}}^{\mathrm{miss}}}\xspace}

%

\newcommand{\ga}{\ensuremath{\gtrsim}}
\newcommand{\la}{\ensuremath{\lesssim}}
\newcommand{\swsq}{\ensuremath{\sin^2\theta_W}\xspace}
\newcommand{\cwsq}{\ensuremath{\cos^2\theta_W}\xspace}
\newcommand{\tanb}{\ensuremath{\tan\beta}\xspace}
\newcommand{\tanbsq}{\ensuremath{\tan^{2}\beta}\xspace}
\newcommand{\sidb}{\ensuremath{\sin 2\beta}\xspace}
\newcommand{\alpS}{\ensuremath{\alpha_S}\xspace}
\newcommand{\alpt}{\ensuremath{\tilde{\alpha}}\xspace}

\newcommand{\QL}{\ensuremath{Q_L}\xspace}
\newcommand{\sQ}{\ensuremath{\tilde{Q}}\xspace}
\newcommand{\sQL}{\ensuremath{\tilde{Q}_L}\xspace}
\newcommand{\ULC}{\ensuremath{U_L^C}\xspace}
\newcommand{\sUC}{\ensuremath{\tilde{U}^C}\xspace}
\newcommand{\sULC}{\ensuremath{\tilde{U}_L^C}\xspace}
\newcommand{\DLC}{\ensuremath{D_L^C}\xspace}
\newcommand{\sDC}{\ensuremath{\tilde{D}^C}\xspace}
\newcommand{\sDLC}{\ensuremath{\tilde{D}_L^C}\xspace}
\newcommand{\LL}{\ensuremath{L_L}\xspace}
\newcommand{\sL}{\ensuremath{\tilde{L}}\xspace}
\newcommand{\sLL}{\ensuremath{\tilde{L}_L}\xspace}
\newcommand{\ELC}{\ensuremath{E_L^C}\xspace}
\newcommand{\sEC}{\ensuremath{\tilde{E}^C}\xspace}
\newcommand{\sELC}{\ensuremath{\tilde{E}_L^C}\xspace}
\newcommand{\sEL}{\ensuremath{\tilde{E}_L}\xspace}
\newcommand{\sER}{\ensuremath{\tilde{E}_R}\xspace}
\newcommand{\sFer}{\ensuremath{\tilde{f}}\xspace}
\newcommand{\sQua}{\ensuremath{\tilde{q}}\xspace}
\newcommand{\sUp}{\ensuremath{\tilde{u}}\xspace}
\newcommand{\suL}{\ensuremath{\tilde{u}_L}\xspace}
\newcommand{\suR}{\ensuremath{\tilde{u}_R}\xspace}
\newcommand{\sDw}{\ensuremath{\tilde{d}}\xspace}
\newcommand{\sdL}{\ensuremath{\tilde{d}_L}\xspace}
\newcommand{\sdR}{\ensuremath{\tilde{d}_R}\xspace}
\newcommand{\sTop}{\ensuremath{\tilde{t}}\xspace}
\newcommand{\stL}{\ensuremath{\tilde{t}_L}\xspace}
\newcommand{\stR}{\ensuremath{\tilde{t}_R}\xspace}
\newcommand{\stone}{\ensuremath{\tilde{t}_1}\xspace}
\newcommand{\sttwo}{\ensuremath{\tilde{t}_2}\xspace}
\newcommand{\sBot}{\ensuremath{\tilde{b}}\xspace}
\newcommand{\sbL}{\ensuremath{\tilde{b}_L}\xspace}
\newcommand{\sbR}{\ensuremath{\tilde{b}_R}\xspace}
\newcommand{\sbone}{\ensuremath{\tilde{b}_1}\xspace}
\newcommand{\sbtwo}{\ensuremath{\tilde{b}_2}\xspace}
\newcommand{\sLep}{\ensuremath{\tilde{l}}\xspace}
\newcommand{\sLepC}{\ensuremath{\tilde{l}^C}\xspace}
\newcommand{\sEl}{\ensuremath{\tilde{e}}\xspace}
\newcommand{\sElC}{\ensuremath{\tilde{e}^C}\xspace}
\newcommand{\seL}{\ensuremath{\tilde{e}_L}\xspace}
\newcommand{\seR}{\ensuremath{\tilde{e}_R}\xspace}
\newcommand{\snL}{\ensuremath{\tilde{\nu}_L}\xspace}
\newcommand{\sMu}{\ensuremath{\tilde{\mu}}\xspace}
\newcommand{\sNu}{\ensuremath{\tilde{\nu}}\xspace}
\newcommand{\sTau}{\ensuremath{\tilde{\tau}}\xspace}
\newcommand{\Glu}{\ensuremath{g}\xspace}
\newcommand{\sGlu}{\ensuremath{\tilde{g}}\xspace}
\newcommand{\Wpm}{\ensuremath{W^{\pm}}\xspace}
\newcommand{\sWpm}{\ensuremath{\tilde{W}^{\pm}}\xspace}
\newcommand{\Wz}{\ensuremath{W^{0}}\xspace}
\newcommand{\sWz}{\ensuremath{\tilde{W}^{0}}\xspace}
\newcommand{\sWino}{\ensuremath{\tilde{W}}\xspace}
\newcommand{\Bz}{\ensuremath{B^{0}}\xspace}
\newcommand{\sBz}{\ensuremath{\tilde{B}^{0}}\xspace}
\newcommand{\sBino}{\ensuremath{\tilde{B}}\xspace}
\newcommand{\Zz}{\ensuremath{Z^{0}}\xspace}
\newcommand{\sZino}{\ensuremath{\tilde{Z}^{0}}\xspace}
\newcommand{\sGam}{\ensuremath{\tilde{\gamma}}\xspace}
\newcommand{\chiz}{\ensuremath{\tilde{\chi}^{0}}\xspace}
\newcommand{\chip}{\ensuremath{\tilde{\chi}^{+}}\xspace}
\newcommand{\chim}{\ensuremath{\tilde{\chi}^{-}}\xspace}
\newcommand{\chipm}{\ensuremath{\tilde{\chi}^{\pm}}\xspace}
\newcommand{\Hone}{\ensuremath{H_{d}}\xspace}
\newcommand{\sHone}{\ensuremath{\tilde{H}_{d}}\xspace}
\newcommand{\Htwo}{\ensuremath{H_{u}}\xspace}
\newcommand{\sHtwo}{\ensuremath{\tilde{H}_{u}}\xspace}
\newcommand{\sHig}{\ensuremath{\tilde{H}}\xspace}
\newcommand{\sHa}{\ensuremath{\tilde{H}_{a}}\xspace}
\newcommand{\sHb}{\ensuremath{\tilde{H}_{b}}\xspace}
\newcommand{\sHpm}{\ensuremath{\tilde{H}^{\pm}}\xspace}
\newcommand{\hz}{\ensuremath{h^{0}}\xspace}
\newcommand{\Hz}{\ensuremath{H^{0}}\xspace}
\newcommand{\Az}{\ensuremath{A^{0}}\xspace}
\newcommand{\Hpm}{\ensuremath{H^{\pm}}\xspace}
\newcommand{\sGra}{\ensuremath{\tilde{G}}\xspace}
\newcommand{\mtil}{\ensuremath{\tilde{m}}\xspace}
\newcommand{\rpv}{\ensuremath{\rlap{\kern.2em/}R}\xspace}
\newcommand{\LLE}{\ensuremath{LL\bar{E}}\xspace}
\newcommand{\LQD}{\ensuremath{LQ\bar{D}}\xspace}
\newcommand{\UDD}{\ensuremath{\overline{UDD}}\xspace}
\newcommand{\Lam}{\ensuremath{\lambda}\xspace}
\newcommand{\Lamp}{\ensuremath{\lambda'}\xspace}
\newcommand{\Lampp}{\ensuremath{\lambda''}\xspace}
\newcommand{\spinbd}[2]{\ensuremath{\bar{#1}_{\dot{#2}}}\xspace}

\newcommand{\MD}{\ensuremath{{M_\mathrm{D}}}\xspace}
\newcommand{\Mpl}{\ensuremath{{M_\mathrm{Pl}}}\xspace}
\newcommand{\Rinv} {\ensuremath{{R}^{-1}}\xspace}

%
%
\hyphenation{en-viron-men-tal}

%% file: CFT-09-012-authorlist.tex
\textbf{Yerevan Physics Institute,  Yerevan,  Armenia}\\*[0pt]
S.~Chatrchyan, V.~Khachatryan, A.M.~Sirunyan
\vskip\cmsinstskip
\textbf{Institut f\"{u}r Hochenergiephysik der OeAW,  Wien,  Austria}\\*[0pt]
W.~Adam, B.~Arnold, H.~Bergauer, T.~Bergauer, M.~Dragicevic, M.~Eichberger, J.~Er\"{o}
% ...
\vskip\cmsinstskip
% ...
